\documentclass[aps,prb,twocolumn,notitlepage,superscriptaddress,floatfix]{revtex4-1}
\usepackage{epstopdf}
\usepackage{graphicx}
\usepackage{dcolumn}
\usepackage{bm}
\usepackage{bbm}
\usepackage{color}
\usepackage{xcolor, soul}
\sethlcolor{green}
\usepackage{amsfonts}
\usepackage{amsmath}
\usepackage{mdframed}
\usepackage[normalem]{ulem}
\usepackage{mathrsfs}   
\usepackage[percent]{overpic}
\usepackage[colorlinks=true,citecolor=blue]{hyperref}
\hypersetup{colorlinks=true,citecolor=blue,linkcolor=blue,urlcolor=blue}
\begin{document}
\title{Theory of plasmonic effects in nonlinear optics: the case of graphene}
\author{Habib Rostami}
\email{Habib.Rostami@iit.it}
\affiliation{Istituto Italiano di Tecnologia, Graphene Labs, Via Morego 30, I-16163 Genova,~Italy}
\author{Mikhail I. Katsnelson}
\affiliation{Radboud University, Institute for Molecules and Materials, NL-6525 AJ Nijmegen,~The Netherlands}
\author{Marco Polini}
\affiliation{Istituto Italiano di Tecnologia, Graphene Labs, Via Morego 30, I-16163 Genova,~Italy}
\begin{abstract}
We develop a microscopic large-$N$ theory of electron-electron interaction corrections to multi-legged Feynman diagrams describing second- and third-order nonlinear response functions. Our theory, which reduces to the well-known random phase approximation in the linear-response limit, is completely general and is useful to understand all second- and third-order nonlinear effects, including harmonic generation, wave mixing, and photon drag. We apply our theoretical framework to the case of graphene, by carrying out microscopic calculations of the second- and third-order nonlinear response functions of an interacting two-dimensional (2D) gas of massless Dirac fermions. We compare our results with recent measurements, where all-optical launching of graphene plasmons has been achieved by virtue of the finiteness of the quasi-homogeneous second-order nonlinear response of this inversion-symmetric 2D material.
\end{abstract}
\maketitle
\section{Introduction}
Recent years have witnessed momentous interest~\cite{koppens_nanolett_2011,grigorenko_naturephoton_2012,basov_rmp_2014,low_acsnano_2014,abajo_acsphoton_2014} in the collective density oscillations of a doped graphene sheet, the so-called Dirac plasmons.  The reason is partly related to the fact that the propagation of graphene plasmons has been {\it directly} imaged in real space by utilizing scattering-type scanning near-field optical microscopy~\cite{fei_nature_2012,chen_nature_2012}. In a series of pioneering experiments in the mid-infrared spectral range, Fei {\it et al.}~\cite{fei_nature_2012} and Chen {\it et al.}~\cite{chen_nature_2012} demonstrated that the plasmon wavelength $\lambda_{\rm p}$ can be $\approx 40$-$60$ times smaller than the free-space excitation wavelength $\lambda_0 =2\pi c/\omega$, allowing an extreme concentration of electromagnetic energy, and that Dirac plasmon properties are easily {\it gate tunable}. 

Importantly, these figures of merit have been dramatically improved by utilizing van der Waals stacks~\cite{geim_nature_2013} comprising graphene encapsulated in boron nitride crystals~\cite{mayorov_nanolett_2011,wang_science_2013,bandurin_science_2016}. Mid-infrared plasmons in encapsulated graphene display~\cite{woessner_naturemater_2015} ultra-large field confinement (i.e.~$\lambda_{\rm p} \approx \lambda_{0}/150$), small group velocity, and a remarkably long lifetime, $\gtrsim 500~{\rm fs}$. In the Terahertz spectral range, acoustic plasmons with $\lambda_{\rm p} \approx \lambda_{0}/66$ and a similar lifetime have been recently observed in hBN/graphene/hBN heterojunctions including a nearby metal gate~\cite{alonsogonzalez_naturenano_2016}.

Substantial theoretical efforts have also been devoted to understanding the nonlinear optical properties of graphene~\cite{jafari_jpcm_2012,mikhailov_prb_2014,cheng_njp_2014,wehling_prb_2015,cheng_prb_2015,Habib_prb_2016,mikhailov_prb_2016,mikhailov_arxiv_2016,savostianova_arxiv_2016,cheng_arxiv_2016}. Experimentally, Hendry et al.~\cite{hendry_prl_2010} demonstrated that the third-order optical susceptibility of graphene is remarkably large ($\approx 1.4\times10^{-15} ~{\rm m}^2/{\rm V}^2$) and only weakly dependent on wavelength in the near-infrared frequency range. Third-harmonic generation (THG) from mechanically exfoliated graphene sheets has been measured by Kumar et al.~\cite{kumar_prb_2013} who extracted a value of the third-order susceptibility on the order of $10^{-16}~{\rm m}^2/{\rm V}^2$ for an incident photon energy $\hbar\omega=0.72~{\rm eV}$. Finally, Hong et al.~\cite{hong_prx_2013} reported strong THG in graphene grown by chemical vapor deposition, in the situation in which the incident photon energy $\hbar\omega=1.57~{\rm eV}$ is in three-photon resonance with the exciton-shifted van Hove singularity.

Since plasmons enable the concentration of electromagnetic energy into extremely small volumes, many groups have theoretically studied the interplay between plasmons and the nonlinear optical properties of graphene and its nanostructures~\cite{Mikhailov_prb_2011,Mikhailov_prl_2014,Yao_prl_2014,Manzoni_njp_2015,Tokman_prb_2016,cox_acsnano_2016,Constant_arXiv_2016}. An all-optical plasmon coupling scheme, which takes advantage of the intrinsic nonlinear optical response of graphene, has been implemented experimentally~\cite{Constant_np_2016}. Free-space, visible light pulses were used by the authors of Ref.~\onlinecite{Constant_np_2016} to launch Dirac plasmons in graphene. Difference-frequency wave mixing (see below) enabled the achievement of the energy- and wavevector-matching conditions. By carefully controlling the phase matching conditions, they also showed that one can excite Dirac plasmons with a definite wavevector and direction across a large frequency range, with an estimated efficiency approaching $10^{-5}$.

In this Article, we present a formal theory of second- and third-order nonlinearities, which treats quantum effects, intra- and inter-band contributions, and electron-electron interactions on equal footing. Our theory starts from an equilibrium Matsubara approach and related Feynman diagrams for the non-interacting nonlinear susceptibilities~\cite{Habib_prb_2016} and includes electron-electron interactions via a large-$N$ approach~\cite{Coleman}. Here, $N$ refers to the number of fermion flavors. In this approximation, only diagrams with the largest number of fermion loops are kept~\cite{Coleman}, with the idea that each fermion loop (bubble) carries a factor $N$. Our large-$N$ theory reduces to the ordinary Bohm-Pines random phase approximation (RPA)~\cite{Pines_and_Nozieres,Giuliani_and_Vignale} in the case of linear response theory. This Article therefore naturally generalizes RPA theory to the case of nonlinear response functions, capturing screening and plasmons. 

While our approach is completely general, we carry out detailed microscopic calculations for the case of a system of two-dimensional (2D) massless Dirac fermions~\cite{Katsnelson,kotov_rmp_2012} interacting via long-range Coulomb interactions.  Large-$N$ theories are known to work very well for weakly correlated materials, like graphene, in which long-range Coulomb interactions (rather than on-site Hubbard-type interactions) play a major role, while they fail to describe e.g.~excitonic effects in semiconductors. In this case, vertex corrections need to be taken into account. This is well beyond the scope of the present Article and is left for future works.

\begin{figure*}[t]
\includegraphics[width=1.0\linewidth]{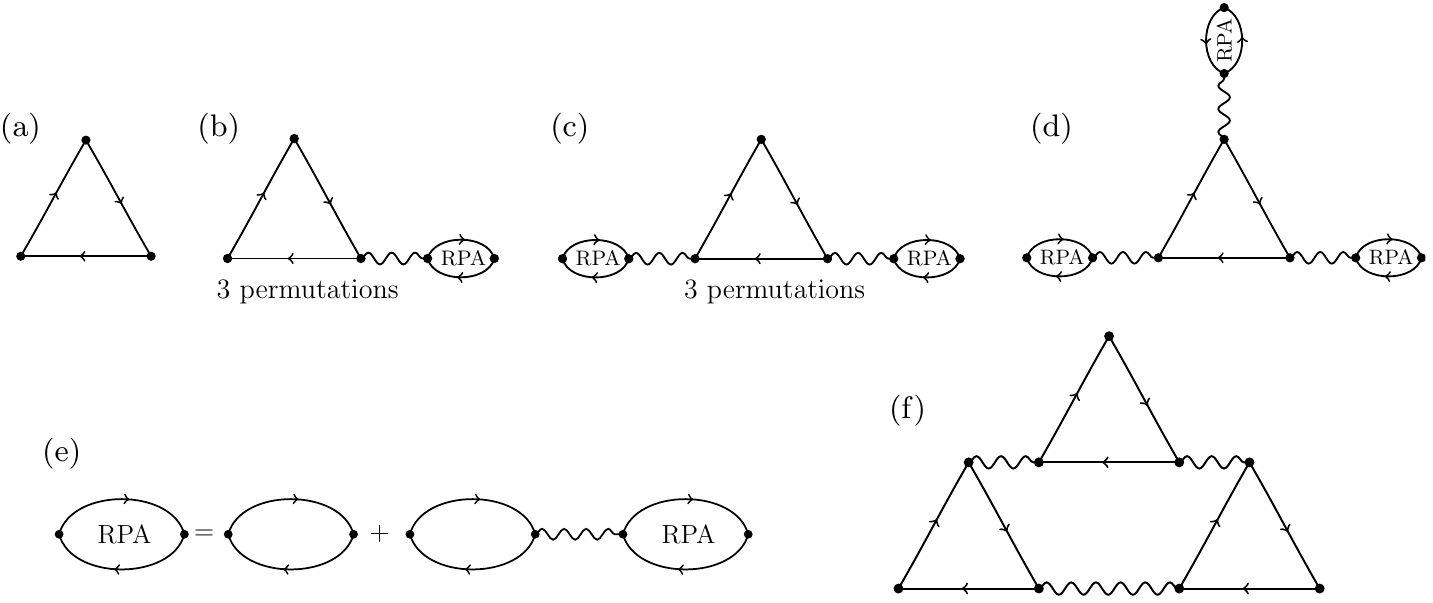}
\caption{Large-$N$ theory for the second-order density response function, $\chi^{(2)}$. 
(a) The Feynman diagram for the non-interacting second-order density response function, $\chi^{(2)}_{0}$.
(b)-(d) Feynman diagrams for the second-order density response function in the large-$N$ approximation.
(e) Infinite series of Feynman diagrams for the linear density response function: Dyson equation in the large-$N$ limit. 
(f) Example of a Feynman diagram for the second-order response function, which is {\it excluded} from the large-$N$ approximation. Solid lines stand for non-interacting Green's functions, solid circles represent density vertices, and wavy lines denote the electron-electron interaction $v_{\bm q}$.\label{fig:chi2_RPA}}
\end{figure*}

Our Article is organized as following. 
In Sect.~\ref{sect:second-order} we present our large-$N$ theory of the second-order susceptibility, while the case of the third-order response is analyzed in Sect.~\ref{sect:third-order}. In Sect.~\ref{sect:small_q_limit} we detail the derivation of long-wavelength expressions for the second- and third-order density response functions, which are extremely useful to understand nonlinear optics experiments. Symmetry considerations that apply to the case of homogeneous and isotropic 2D systems are summarized in Sect.~\ref{sect:symmetries}. In Sect.~\ref{sect:second-order-Dirac} we present explicit analytical and numerical calculations of the second-order conductivity of a 2D system of massless Dirac fermions. For the sake of clarity, the case of harmonic generation and sum/difference wave mixing are independently analyzed in Sects.~\ref{sect:harmonic} and~\ref{sect:sum_and_difference}, respectively. A comparison between our theory and available experimental results~\cite{Constant_np_2016} is reported in Sect.~\ref{subsection:comparison-with-experimental-results}. A summary of our main results and a brief set of conclusions are reported in Sect.~\ref{sect:summary}. Three appendices report a wealth of technical details. In particular, in Appendix~\ref{app:TDH}, we show that the main formal results of our Article, Eqs.~(\ref{eq:chi2_RPA}), (\ref{eq:chi3_RPAa}), and~ (\ref{eq:chi3_RPAb}), can also be independently derived by using the time-dependent Hartree approximation~\cite{Giuliani_and_Vignale}. This derivation highlights a pathway to transcend the large-$N$ approximation, by suggesting a straightforward approach to include exchange and correlation effects in the spirit of density functional theory~\cite{Giuliani_and_Vignale}.

\section{Second-order density response in the large-$N$ limit}
\label{sect:second-order}

We start by considering the bare second-order density response function. This is diagrammatically represented in Fig.~\ref{fig:chi2_RPA}(a). In this diagram (usually termed ``triangle'' diagram), solid lines are non-interacting Green's functions while filled circles represent density vertices~\cite{Habib_prb_2016}. 

In the large-$N$ approximation~\cite{Coleman}, electron-electron interactions are captured by the diagrams reported in Figs.~\ref{fig:chi2_RPA}(b)-(d). In Fig.~\ref{fig:chi2_RPA}(b), all the three density vertices of the bare diagram are dressed by the infinite RPA series of bubble diagrams shown in Fig.~\ref{fig:chi2_RPA}(e). Similarly, in Fig.~\ref{fig:chi2_RPA}(c) and (d), only two vertices (one vertex) are (is) dressed by the RPA series of bubble diagrams. 

The logic of keeping only the diagrams in Fig.~\ref{fig:chi2_RPA}(b)-(d) is the following. In the large-$N$ approximation, only diagrams that dominate in the limit $N \to \infty$ are retained, where $N$ stands for the number of fermion flavors ($N=4$, for example, in graphene). As discussed in Ref.~\onlinecite{Coleman}, each fermion loop (i.e.~bubble) brings a factor $N$, while each electron-electron interaction line brings a factor $1/N$. Therefore, a diagram with $n_{\rm b}$ bubbles and $n_{\rm v}$ electron-electron interaction lines scales like $N^{n}$, where $n=n_{\rm b}-n_{\rm v}$. In the limit $N \to \infty$, all diagrams with $n \le 0$ are negligible, while diagrams with $n>0$ dominate. Diagrams in Fig.~\ref{fig:chi2_RPA}(b)-(d) have $n = 1$, while the diagram in Fig.~\ref{fig:chi2_RPA}(f) has $n=0$. We therefore conclude that the latter diagram must be discarded in the large-$N$ approximation.

The sum of diagrams in Fig.~\ref{fig:chi2_RPA}(b)-(d) renormalizes the bare second-order response $\chi^{(2)}_{0}$ shown in Fig.~\ref{fig:chi2_RPA}(a).
We find that the second-order nonlinear response function in the large-$N$ limit, which will be denoted by the symbol $\chi^{(2)}$, is given by the following expression:
\begin{equation}\label{eq:chi2_RPA_0}
\chi^{(2)}= \frac{\chi^{(2)}_{0}}{{\cal R}_2}~,
\end{equation} 
where
\begin{eqnarray}\label{eq:R2}
\frac{1}{{\cal R}_2} &\equiv& 
1+\sum_{i=1,2,\Sigma} v_{i} \chi^{(1)}(i)\nonumber\\
&+&  \sum_{i=1,2,\Sigma} \frac{v_{1} \chi^{(1)}(1) v_{2} \chi^{(1)}(2)v_{\Sigma} \chi^{(1)}(\Sigma)}{v_{i} \chi^{(1)}(i)}
\nonumber\\
&+& v_{1} \chi^{(1)}(1) v_{2} \chi^{(1)}(2)v_{\Sigma} \chi^{(1)}(\Sigma)~.
\end{eqnarray}
In Eq.~(\ref{eq:R2}) we have used the following shorthand: $v_i \equiv v_{{\bm q}_i}$ and $\chi^{(1)} \equiv \chi^{(1)}(-{\bm q}_i,{\bm q}_i,-\omega_i,\omega_i)$. 
Here, $v_{\bm q}$ is the 2D Fourier transform of the Coulomb interaction potential and
\begin{equation}
\chi^{(1)}(-{\bm q},{\bm q},-\omega,\omega)  \equiv \frac{\chi^{(1)}_{0}({\bm q},\omega)}{1- v_{\bm q}\chi^{(1)}_{0}({\bm q},\omega)}~,
\end{equation}
is the usual RPA series of bubble diagrams~\cite{Pines_and_Nozieres,Giuliani_and_Vignale}, where $\chi^{(1)}_{0}({\bm q},\omega)$ is the frequency- and wavevector-dependent first-order {\it non-interacting} density response function~\cite{Pines_and_Nozieres,Giuliani_and_Vignale}. Finally, for $i = \Sigma$, we have ${\bm q}_{\Sigma} \equiv {\bm q}_{1} + {\bm q}_2$ and $\omega_{\Sigma} \equiv \omega_{1} + \omega_{2}$.

Carrying out the sums in Eq.~(\ref{eq:R2}), we find 
\begin{equation}\label{eq:R2_explicit}
{\cal R}_2= \epsilon(\Sigma)  \epsilon(2) \epsilon(1)~,
\end{equation}
where $\epsilon(i)$ is a shorthand for
\begin{equation}
\epsilon({\bm q}_i,\omega_i) = 1 - v_{{\bm q}_i}\chi^{(1)}_{0}({\bm q}_i,\omega_i)~, 
\end{equation}
which is the dynamical RPA screening function~\cite{Pines_and_Nozieres,Giuliani_and_Vignale}. 

Therefore, the second-order density-density response function in the large-$N$ limit is given by
\begin{eqnarray}\label{eq:chi2_RPA}
&&\chi^{(2)}(-{\bm q}_\Sigma,{\bm q}_1,{\bm q}_2,-\omega_\Sigma,\omega_1,\omega_2) =
\nonumber \\ &&
 \frac{\chi^{(2)}_{0}(-{\bm q}_\Sigma,{\bm q}_1,{\bm q}_2,-\omega_\Sigma,\omega_1,\omega_2) }
{  \epsilon({\bm q}_\Sigma,\omega_\Sigma)  \epsilon({\bm q}_2,\omega_2)  \epsilon({\bm q}_1,\omega_1)}~.
\end{eqnarray}
In the harmonic case, $({\bm q}_1,\omega_{1}) =({\bm q}_2, \omega_{2})$, Eq.~(\ref{eq:chi2_RPA}) reduces to a result that has been obtained earlier by using a self-consistent density-matrix approach~\cite{Mikhailov_prb_2011,Mikhailov_prl_2014}. 

\section{Third-order density response in the large-$N$ limit}
\label{sect:third-order}

\begin{figure*}[t]
\includegraphics[width=0.8\linewidth]{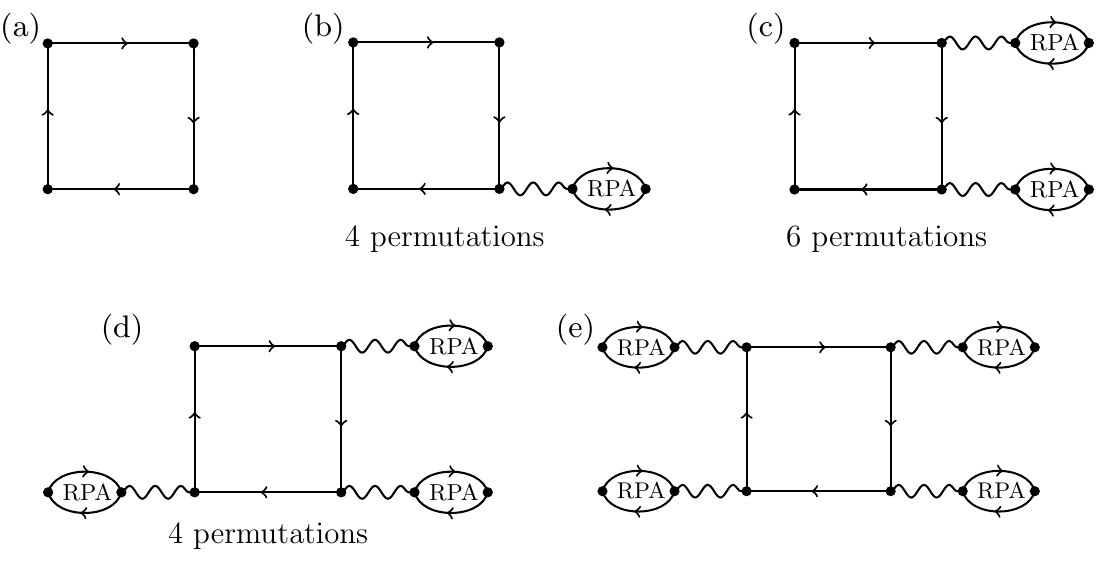}
\caption{First family of large-$N$ diagrams for the third-order density response function. These are obtained by renormalizing the vertices of the non-interacting third-order density response function, shown in panel (a).
(b)-(e) Feynman diagrams for the third-order density response function in the large-$N$ approximation.\label{fig:chi3_RPA}}
\end{figure*}

\begin{figure*}[t]
\includegraphics[width=\linewidth]{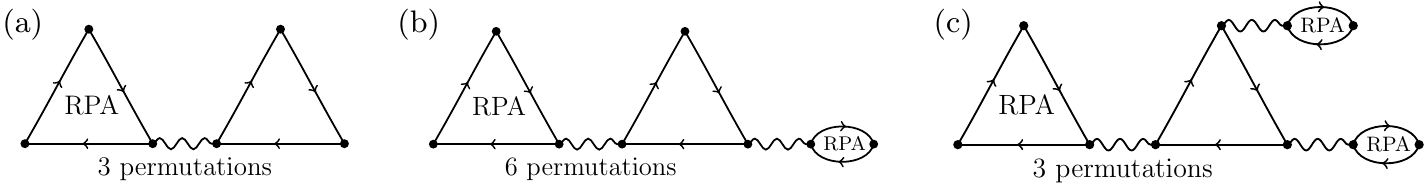}
\caption{Second family of Feynman diagrams for the third-order response function. These are obtained by glueing together two non-interacting second-order diagrams via an electron-electron interaction line. 
\label{fig:chi3_bis_RPA}}
\end{figure*}

In this Section we lay down a large-$N$ theory for the third-order response function. 
In this case, the situation is more subtle. The point is that the bare third-order response function (square diagram) contains {\it four} density vertices, see Fig.~\ref{fig:chi3_RPA}(a). One can therefore create {\it two} families of large-$N$ diagrams that contribute to the third-order response. The first family, which is based on square-type diagrams, is shown in Fig.~\ref{fig:chi3_RPA}.  These large-$N$ series just renormalizes the bare third-order response, as in the case of the second-order response in Fig.~\ref{fig:chi2_RPA}. The second family is topologically distinct and based on triangle-type diagrams, as illustrated in Fig.~\ref{fig:chi3_bis_RPA}. The idea of the second family is that you can create Feynman diagrams with four density vertices by ``glueing'' together two second-order triangular diagrams via an electron-electron interaction line. The sum of the diagrams in the first family will be denoted by the symbol $\chi^{(3)}_{\rm a}$, while the sum of the diagrams in the second family will be denoted by $\chi^{(3)}_{\rm b}$. The full third-order response function in the large-$N$ approximation is given by: $\chi^{(3)}= \chi^{(3)}_{\rm a}+\chi^{(3)}_{\rm b}$.

In analogy with the second-order response function (\ref{eq:chi2_RPA}), the sum of the diagrams in Fig.~\ref{fig:chi3_RPA} can be written as
\begin{equation}\label{eq:sum_type_I}
\chi^{(3)}_{\rm a} = \frac{\chi^{(3)}_0}{{\cal R}_3}~, 
\end{equation}
where
\begin{eqnarray}\label{eq:R3}
&&\frac{1}{{\cal R}_{3}}
\equiv
1+\sum_{i} v_{i} \chi^{(1)}(i) +
\nonumber\\
&&\frac{1}{2}\sum_{i,j}   \frac{v_{1} \chi^{(1)}(1) v_{2} \chi^{(1)}(2) v_{3} \chi^{(1)}(3) v_{\Sigma} \chi^{(1)}(\Sigma)}{v_{i} \chi^{(1)}(i) v_{j} \chi^{(1)}(j)} (1-\delta_{ij})+
\nonumber \\
&&\sum_{i} \frac{v_{1} \chi^{(1)}(1) v_{2} \chi^{(1)}(2) v_{3} \chi^{(1)}(3) v_{\Sigma} \chi^{(1)}(\Sigma)}{v_{i} \chi^{(1)}(i)}+
\nonumber\\
&&v_{1} \chi^{(1)}(1) v_{2} \chi^{(1)}(2)v_{3} \chi^{(1)}(3) v_{\Sigma} \chi^{(1)}(\Sigma)~.
\end{eqnarray}
Here, $i,j= \{1,2,3,\Sigma\}$ and $\delta_{ij}$ is the Kronecker delta.
Carrying out the sums in Eq.~(\ref{eq:R3}), we find
\begin{equation}
{\cal R}_3= \epsilon(\Sigma) \epsilon(3) \epsilon(2) \epsilon(1)
\end{equation}
and
\begin{eqnarray}\label{eq:chi3_RPAa}
&&\chi^{(3)}_{\rm a}(-{\bm q}_\Sigma,{\bm q}_1,{\bm q}_2,{\bm q}_3,-\omega_\Sigma,\omega_1,\omega_2,\omega_3) =
\nonumber\\ &&
 \frac{\chi^{(3)}_{0}(-{\bm q}_\Sigma,{\bm q}_1,{\bm q}_2,{\bm q}_3,-\omega_\Sigma,\omega_1,\omega_2,\omega_3) }
{\epsilon({\bm q}_\Sigma,\omega_\Sigma) \Pi^3_{i=1} \epsilon({\bm q}_i,\omega_i) } ~.
\end{eqnarray}

The situation is quite different for the second family of Feynman diagrams shown in Fig.~\ref{fig:chi3_bis_RPA}. The sum of these diagrams can be written as
\begin{equation}\label{eq:sum_type_II}
\chi^{(3)}_{\rm b} = \sum_{i=1,2,3} \frac{\chi^{(2)}(i) v_i \chi^{(2)}_{0}(i)}{{\cal K}_i}~,
\end{equation}
where
\begin{equation}\label{eq:typeII_ingredient1}
\chi^{(2)}(i) \equiv \chi^{(2)}(-{\bm q}_\Sigma,{\bm q}_i, \widetilde{\bm q}_i,-\omega_\Sigma,\omega_i,\widetilde\omega_i)~,
\end{equation}
\begin{equation}\label{eq:typeII_ingredient2}
\chi^{(2)}_{0}(1) \equiv \chi^{(2)}_{0}(-\widetilde{\bm q}_1,{\bm q}_2,{\bm q}_3,-\widetilde\omega_1,\omega_2,\omega_3)~,
\end{equation}
\begin{equation}\label{eq:typeII_ingredient3}
\chi^{(2)}_{0}(2) \equiv \chi^{(2)}_{0}(-\widetilde{\bm q}_2,{\bm q}_3,{\bm q}_1,-\widetilde\omega_2,\omega_3,\omega_1)~,
\end{equation}
\begin{equation}\label{eq:typeII_ingredient4}
\chi^{(2)}_{0}(3) \equiv \chi^{(2)}_{0}(-\widetilde{\bm q}_3,{\bm q}_1,{\bm q}_2,-\widetilde\omega_3,\omega_1,\omega_2)~,
\end{equation}
and
\begin{eqnarray}\label{eq:typeII_ingredient5}
\frac{1}{{\cal K}_i} &=& 1+ \sum_{j=1,2,3} v_j \chi^{(1)}(j) (1-\delta_{ij})
\nonumber\\
&+&\frac{v_1 \chi^{(1)}(1)v_2 \chi^{(1)}(2)v_3 \chi^{(1)}(3)}{v_i \chi^{(1)}(i)}~.
\end{eqnarray}
In Eqs.~(\ref{eq:typeII_ingredient1})-(\ref{eq:typeII_ingredient4}), $\widetilde {\bm q}_i \equiv {\bm q}_\Sigma-{\bm q}_i$ and $\widetilde\omega_i \equiv \omega_\Sigma-\omega_i$.

From Eq.~(\ref{eq:typeII_ingredient5}), one can show that ${\cal K}_1=  \epsilon(3) \epsilon(2)$, with similar expressions holding for ${\cal K}_2$ and ${\cal K}_3$, provided that suitable cyclic permutations of the $1$,$2$, and $3$ indices are carried out.

After lengthy but straightforward algebra we conclude that
\begin{eqnarray}\label{eq:chi3_RPAb}
&&\chi^{(3)}_{\rm b}(-{\bm q}_\Sigma,{\bm q}_1,{\bm q}_2,{\bm q}_3,-\omega_\Sigma,\omega_1,\omega_2,\omega_3)  =
\nonumber\\ && 
 \sum^3_{i=1} \frac { 
 v_{\rm sc}(\widetilde {\bm q}_i,\widetilde \omega_i)
 \chi^{(2)}_0(-{\bm q}_\Sigma,{\bm q}_i,\widetilde {\bm q}_i,-\omega_\Sigma,\omega_i,\widetilde \omega_i) 
 }{
\epsilon({\bm q}_\Sigma,\omega_\Sigma) 
\Pi^3_{l=1}\epsilon({\bm q}_l,\omega_l)  }  \times
\nonumber\\ && 
 \chi^{(2)}_{0}(-\widetilde {\bm q}_i,{\bm q}_j,{\bm q}_k,-\widetilde \omega_i,\omega_j,\omega_k)~,   
\end{eqnarray}
where $j,k=2,3$ for $i=1$ and so on and so forth, in a cyclic manner, and the dynamically screened interaction is defined by~\cite{Giuliani_and_Vignale}
\begin{equation}
v_{\rm sc}({\bm q},\omega ) \equiv \frac{v_{\bm q}}{\epsilon({\bm q}, \omega)}~.
\end{equation}

An alternative derivation of Eqs.~(\ref{eq:chi2_RPA}), (\ref{eq:chi3_RPAa}), and~(\ref{eq:chi3_RPAb}), which is based on the time-dependent Hartree approximation, is offered in Appendix~\ref{app:TDH}.

\section{Long-wavelength expansion of nonlinear density response functions}
\label{sect:small_q_limit}

In this Section we present a long-wavelength expansion of the nonlinear response functions introduced in the previous Sections. To this end, we take advantage of the gauge invariance principle and introduce nonlinear {\it conductivity} tensors.  

Using gauge invariance~\cite{habib_gaugeinv_2016}, we obtain the following relation between the $n$-th order nonlinear density response function and the corresponding nonlinear conductivity:
\begin{eqnarray}\label{eq:chi-sigma}
\chi^{(n)}   = \frac{(-i)^n}{\omega_\Sigma} \sum_{\ell, \{ \alpha_i \}} 
q_{\Sigma,\ell} \Pi^n_{i=1} q_{i,\alpha_i}
\sigma^{(n)}_{\ell \alpha_1\dots\alpha_n}~,
\end{eqnarray}
where $q_{\Sigma,\ell}$ and $q_{i,\alpha_i}$ are the Cartesian components of the vectors ${\bm q}_{\Sigma}$ and ${\bm q}_i$, respectively. In writing Eq.~(\ref{eq:chi-sigma}) we have dropped for simplicity the argument of the nonlinear functions $\chi^{(n)}$ and $\sigma^{(n)}_{\ell \alpha_1\dots\alpha_n}$: $\chi^{(n)} = \chi^{(n)}(-{\bm q}_\Sigma,{\bm q}_1,\dots,{\bm q}_n,-\omega_\Sigma,\omega_1,\dots,\omega_n)$ and $\sigma^{(n)}_{\ell \alpha_1\dots\alpha_n} = \sigma^{(n)}_{\ell \alpha_1\dots\alpha_n}(-{\bm q}_\Sigma,{\bm q}_1,\dots,{\bm q}_n,-\omega_\Sigma,\omega_1,\dots,\omega_n)$.

Using Eq.~(\ref{eq:chi-sigma}), we can first express the dynamical screening function in terms of the linear-response conductivity tensor:
\begin{equation}
\epsilon({\bm q},\omega) =  1+ i v_{\bm q}  \sum_{\ell \alpha} 
\frac{q_{\ell} q_{\alpha}}{\omega}   \sigma^{(1)}_{\ell \alpha} (-{\bm q},{\bm q} ,-\omega,\omega ) ~.
\end{equation}

Using Eqs.~(\ref{eq:chi2_RPA_0}), (\ref{eq:sum_type_I}), (\ref{eq:sum_type_II}), 
and~(\ref{eq:chi-sigma}), we obtain the following formal relations for the second- and third-order conductivities:
\begin{equation}\label{eq:sigma2_eff}
\sigma^{(2), {\rm ee}}_{\ell \alpha_1\alpha_2}  =  
\frac{\sigma^{(2)}_{\ell \alpha_1\alpha_2}   }{{\cal R}_2}~,
\end{equation}
and
\begin{equation}\label{eq:sigma3_eff}
\sigma^{(3), {\rm ee}}_{\ell \alpha_1\alpha_2\alpha_3}  =  
\frac{\sigma^{(3)}_{\ell \alpha_1\alpha_2\alpha_3}  + {\widetilde \sigma}^{(3)}_{\ell \alpha_1\alpha_2\alpha_3}}{{\cal R}_3}~.
\end{equation}
In Eqs.~(\ref{eq:sigma2_eff}) and~(\ref{eq:sigma3_eff}), $\sigma^{(2), {\rm ee}}_{\ell \alpha_1\alpha_2}$ and $\sigma^{(3), {\rm ee}}_{\ell \alpha_1\alpha_2\alpha_3}$ denote the second- and third-order conductivities of the {\it interacting} electron system, while $\sigma^{(2)}_{\ell \alpha_1\alpha_2}$ and $\sigma^{(3)}_{\ell \alpha_1\alpha_2\alpha_3}$ denote their non-interacting counterparts. 

Also, in Eq.~(\ref{eq:sigma3_eff}) we have introduced
\begin{widetext}
\begin{eqnarray}\label{eq:tilde_sigma3}
\widetilde\sigma^{(3)}_{\ell \alpha_1\alpha_2\alpha_3}  (-{\bm q}_\Sigma,{\bm q}_1,{\bm q}_2,{\bm q}_3,-\omega_\Sigma,\omega_1,\omega_2,\omega_3)
&=& -i \sum^3_{i=1} 
\Bigg\{ \frac{v_{\rm sc}(\widetilde {\bm q}_i,\widetilde\omega_{i})}{\widetilde\omega_{i}}
 \sum_{\beta, \beta'} {\widetilde q}_{i,\beta} {\widetilde q}_{i,\beta'}
\sigma^{(2)}_{\ell \alpha_1\beta} (-{\bm q}_\Sigma, {\bm q}_i, \widetilde {\bm q}_i, -\omega_\Sigma, \omega_i, \widetilde \omega_i)  
\nonumber\\ &\times&
 \sigma^{(2)}_{\beta' \alpha_2\alpha_3}(-\widetilde {\bm q}_i ,{\bm q}_j,{\bm q}_k,-\widetilde \omega_i,\omega_j,\omega_k) 
 \Bigg \}~.
\end{eqnarray}
\end{widetext} 
The contribution denoted by the symbol ${\widetilde \sigma}^{(3)}_{\ell \alpha_1\alpha_2\alpha_3}$ stems from the family of diagrams shown in Fig.~\ref{fig:chi3_bis_RPA}. As we have discussed earlier, a similar contribution does not exist in the case of the second-order response---cf.~Eq.~(\ref{eq:sigma2_eff}). Physically, ${\widetilde \sigma}^{(3)}_{\ell \alpha_1\alpha_2\alpha_3}$ represents an interaction-induced {\it nonlocal} contribution to the third-order conductivity. A caveat is now in order. Some care must be exercised when adding $\sigma^{(3)}_{\ell \alpha_1\alpha_2\alpha_3}$ and ${\widetilde \sigma}^{(3)}_{\ell \alpha_1\alpha_2\alpha_3}$ in Eq.~(\ref{eq:sigma3_eff}). In principle, indeed, one should expand $\sigma^{(3)}_{\ell \alpha_1\alpha_2\alpha_3}$ in powers of wavevectors in the long-wavelength limit, up to the same order that appears in Eq.~(\ref{eq:tilde_sigma3}). This calculation is very cumbersome and will be not carried out in this work. The numerical results in Fig.~\ref{fig:plasmonic_shg_thg}(c) below have been calculated by neglecting nonlocal corrections to $\sigma^{(3)}_{\ell \alpha_1\alpha_2\alpha_3}$.

We are now in the position to expand the nonlinear density response functions in the long-wavelength limit. To this end, we just need to expand the conductivity tensors $\sigma^{(n)}_{\ell \alpha_1\dots\alpha_n}$, keeping the leading contributions. We also need to specify the functional dependence of $v_{\bm q}$ on ${\bm q}$. For long-range Coulomb interactions in a {\it free-standing} graphene sheet, the 2D Fourier transform of the Coulomb potential is given by $v_{\bm q} = 1/(2\epsilon_{0} q)$, with $q=|{\bm q}|$ and $\epsilon_{0}$ the vacuum permittivity. In the long-wavelength $q/k_{\rm F} \ll 1$ limit the RPA dynamical screening function can be expanded as
\begin{equation}\label{eq:expansion_dynamical_screening}
\epsilon({\bm q},\omega) = 1 + i  \frac{ q}{ 2\epsilon_0 \omega}  \sigma^{(1)}_{\rm L}(\omega) + \dots~,
\end{equation}
where we have introduced the longitudinal linear conductivity
\begin{equation}
\sigma^{(1)}_{\rm L}(\omega) \equiv \sum_{\ell, \alpha} \frac{q_{\ell} q_{\alpha}}{q^2}   \sigma^{(1)}_{\ell \alpha} ({\bm 0},{\bm 0} ,-\omega,\omega)~.
\end{equation}
In Eq.~(\ref{eq:expansion_dynamical_screening}) and below, ``$\dots$'' denote higher-order corrections, which vanish faster that the leading term in the long-wavelength limit. 

Similarly, the long-wavelength expansion of the second-order density response function requires an expansion of the second-order conductivity up to linear order in $q_i$:
\begin{eqnarray}\label{eq:sigma2_small_q}
&&\sigma^{(2)}_{\ell \alpha_1\alpha_2} (-{\bm q}_\Sigma,{\bm q}_1,{\bm q}_2,-\omega_\Sigma,\omega_1,\omega_2 )
=
\sigma^{(2)}_{\ell \alpha_1\alpha_2} (-\omega_\Sigma;\omega_1,\omega_2 ) 
\nonumber \\ 
&&+\sum_{i=1,2}\sum_{\beta}  q_{i,\beta} d^{(2)}_{\ell \alpha_1\alpha_2 \beta, i} (-\omega_\Sigma;\omega_1,\omega_2 ) +\dots~,
\end{eqnarray} 
\newline
where the zeroth-order term, $\sigma^{(2)}_{\ell \alpha_1\alpha_2} (-\omega_\Sigma;\omega_1,\omega_2 )\equiv\sigma^{(2)}_{\ell \alpha_1\alpha_2} ({\bm 0},{\bm 0},{\bm 0},-\omega_\Sigma,\omega_1,\omega_2 )$, is the second-order optical conductivity, while its {\it dipole} is defined by
\begin{equation}
d^{(2)}_{\ell \alpha_1\alpha_2 \beta, i} (-\omega_\Sigma;\omega_1,\omega_2 ) \equiv
 \frac{\partial \sigma^{(2)}_{\ell \alpha_1\alpha_2}  }{\partial q_{i,\beta}}\Big |_{\{{\bm q}_1,{\bm q}_2\} \to 0}~.
\end{equation}
Using Eqs.~(\ref{eq:chi2_RPA}), (\ref{eq:chi-sigma}) and~(\ref{eq:sigma2_small_q}), we can rewrite the large-$N$ second-order density response in the long-wavelength limit as
\begin{eqnarray}\label{eq:chi2_RPA_small_q}
&&\chi^{(2)}(-{\bm q}_\Sigma,{\bm q}_1,{\bm q}_2,-\omega_\Sigma,\omega_1,\omega_2) =
  (-i)^2 \frac{q_1 q_2 q_\Sigma}{\omega_\Sigma} 
 \frac{1}{{\cal R}_2 }
 \nonumber \\&&
\times
\left [\sigma^{(2)}_{\rm L}(-\omega_\Sigma; \omega_1,\omega_2) 
+ \sum_i q_i d^{(2)}_{{\rm L}, i}(-\omega_\Sigma;\omega_1,\omega_2) \right]+\dots\nonumber\\
\end{eqnarray}
where 
\begin{eqnarray}
\sigma^{(2)}_{\rm L}(-\omega_\Sigma;\omega_1,\omega_2 )  \equiv &&\sum_{\ell, \alpha_1, \alpha_2} \frac{q_{1,\alpha_1} q_{2,\alpha_2} q_{\Sigma,\ell} }{q_1 q_2 q_\Sigma} \times
\nonumber \\
&&\sigma^{(2)}_{\ell, \alpha_1, \alpha_2}(-\omega_\Sigma;\omega_1,\omega_2 )~,
\end{eqnarray}
and
\begin{eqnarray}
d^{(2)}_{{\rm L},i}(-\omega_\Sigma;\omega_1,\omega_2 )  \equiv&& \sum_{\ell, \alpha_1, \alpha_2, \beta} \frac{q_{1,\alpha_1} q_{2,\alpha_2} q_{\Sigma,\ell} q_{i,\beta}}{q_1 q_2 q_\Sigma q_i } \times
\nonumber \\
&&d^{(2)}_{\ell \alpha_1\alpha_2 \beta,i} (-\omega_\Sigma;\omega_1,\omega_2 )~.
\end{eqnarray}
In Eq.~(\ref{eq:chi2_RPA_small_q}), we have introduced the following long-wavelength expansion of the ${\cal R}_n$ factors:
\begin{eqnarray}\label{eq:R_n}
{\cal R}_{n} &=& 
\left [1 +i  \frac{ q_\Sigma}{ 2\epsilon_0\omega_\Sigma}  \sigma^{(1)}_{\rm L}(\omega_\Sigma)  \right ]
\Pi^n_{j=1}\left [1 +i  \frac{ q_j}{2\epsilon_0 \omega_j}  \sigma^{(1)}_{\rm L}(\omega_j)\right]
\nonumber \\&+& \dots
\end{eqnarray}
with $n=2,3, \dots$. 

Similarly, we can expand the third-order nonlinear density response functions. In the long-wavelength limit Eq.~(\ref{eq:sum_type_I}) reduces to
\begin{eqnarray} 
&&\chi^{(3)}_{\rm a}(-{\bm q}_\Sigma,{\bm q}_1,{\bm q}_2,{\bm q}_3,-\omega_\Sigma,\omega_1,\omega_2,\omega_3) =
  (-i)^3 \frac{q_1 q_2 q_3 q_\Sigma}{\omega_\Sigma} 
  \nonumber \\ && \times   \frac{1}{{\cal R}_3}
\sigma^{(3)}_{\rm L}(-\omega_\Sigma;\omega_1,\omega_2,\omega_3) + \dots
\end{eqnarray}
where the longitudinal third-order optical conductivity is given by
\begin{eqnarray}\label{eq:s3L}
\sigma^{(3)}_{\rm L}(-\omega_\Sigma; \omega_1,\omega_2,\omega_3) &=& \sum_{\ell, \{\alpha_i \} } \frac{q_{1,\alpha_1} q_{2,\alpha_2} q_{3,\alpha_3} q_{\Sigma,\ell} }{q_1 q_2 q_3 q_\Sigma} \times
\nonumber \\&~&
\hspace{-3mm}
\sigma^{(3)}_{\ell \alpha_1\alpha_2 \alpha_3} (-\omega_\Sigma; \omega_1,\omega_2,\omega_3 )~.
\end{eqnarray}
In the same limit, Eq.~(\ref{eq:sum_type_II}) reduces to
\begin{eqnarray} 
&& \chi^{(3)}_{\rm b}(-{\bm q}_\Sigma,{\bm q}_1,{\bm q}_2,{\bm q}_3,-\omega_\Sigma,\omega_1,\omega_2,\omega_3) 
=
 \frac{ q_1 q_2 q_3 q_\Sigma}{\omega_\Sigma}  \frac{ (-i)^4}{2\epsilon_0 {\cal R}_3}
\nonumber \\&&
 \sum^3_{i=1} \frac{{\widetilde q_i }/{\widetilde\omega_i} }{ \left [1 -i  \frac{ \widetilde q_i }{2\epsilon_0\widetilde\omega_i}  \sigma^{(1)}_{\rm L}(\widetilde\omega_i)  \right ] }
\Big [ \sigma^{(2)}_{\rm L}(-\omega_\Sigma;\omega_i,\widetilde\omega_i)
+
\nonumber\\ &&
q_i d^{(2)}_{\rm L,i}(-\omega_\Sigma;\omega_i,\widetilde\omega_i)
+ \widetilde q_i d^{(2)}_{\rm L,2}(-\omega_\Sigma;\omega_i,\widetilde\omega_i) 
\Big ] 
\Big[ \sigma^{(2)}_{\rm L}(-\widetilde\omega_i;\omega_j,\omega_k)
\nonumber\\ &&
 +q_j d^{(2)}_{\rm L,1}(-\widetilde\omega_i;\omega_j,\omega_k)
+ q_k d^{(2)}_{\rm L,2}(-\widetilde\omega_i;\omega_j,\omega_k)
\Big]+\dots.
\end{eqnarray}
Notice again that $j,k=2,3$ for $i=1$, and so on and so forth, in a cyclic way.
\section{Symmetry considerations for homogeneous and isotropic 2D systems}
\label{sect:symmetries}

In a homogeneous and isotropic 2D system, the properties of the nonlinear conductivity tensors are highly constrained by rotational, translational, and inversion symmetries. 

We start by recalling that, due to mirror $x \to -x$ ($y \to -y$) symmetry, all elements of the third-order conductivity with an odd number of $x$ ($y$) Cartesian indices are identically zero. Full rotational symmetry implies 
\begin{equation}\label{eq:sym3_1}
\sigma^{(3)}_{xxxx} = \sigma^{(3)}_{xxyy} +  \sigma^{(3)}_{xyxy} + \sigma^{(3)}_{xyyx}~.
\end{equation}
Moreover, mirror symmetry with respect to diagonal in the ${\hat {\bm x}}$-${\hat {\bm y}}$ plane provides an exchange symmetry between $x$ and $y$ Cartesian indices: we therefore have
\begin{eqnarray} \label{eq:sym3_2}
&&\sigma^{(3)}_{yyyy} =\sigma^{(3)}_{xxxx},~~~  \sigma^{(3)}_{yyxx}= \sigma^{(3)}_{xxyy}
\nonumber\\
&&\sigma^{(3)}_{yxyx}= \sigma^{(3)}_{xyxy},~~~ \sigma^{(3)}_{yxxy}= \sigma^{(3)}_{xyyx}~.
\end{eqnarray}
By using Eqs.~(\ref{eq:s3L}), (\ref{eq:sym3_1}), and~(\ref{eq:sym3_2}) and ${\bm q}_i/q_i=\cos(\theta_i) \hat {\bm  x}+\sin(\theta_i) \hat {\bm  y}$ for ${\bm q}_i \neq {\bm 0}$, we get 
\begin{eqnarray} \label{eq:sigma3L}
\sigma^{(3)}_{\rm L} &=&
\frac{\sigma^{(3)}_{xxyy} +\sigma^{(3)}_{xyxy}}{2} \cos(\theta_1+\theta_2-\theta_3-\theta_\Sigma)
\nonumber \\ &+&
\frac{\sigma^{(3)}_{xxyy} +\sigma^{(3)}_{xyyx}}{2} \cos(\theta_1+\theta_3-\theta_2-\theta_\Sigma)
\nonumber \\ &+&
\frac{\sigma^{(3)}_{xyxy} +\sigma^{(3)}_{xyyx}}{2} \cos(\theta_2+\theta_3-\theta_1-\theta_\Sigma)~.
\end{eqnarray}
Here, $\theta_\Sigma$ is the azimuthal angle of ${\bm q}_{\Sigma} \neq {\bm 0}$, i.e.
\begin{equation}
\cos(\theta_\Sigma-\theta_1) = \frac{q_1+q_2 \cos(\theta_{21})+q_3 \cos(\theta_{31})}
{\sqrt{\sum^3_{i=1} q^2_i+2\sum_{i > j}q_i q_j \cos(\theta_{i j})}}
\end{equation}
with $\theta_{ij} \equiv \theta_i-\theta_j$. All conductivity tensor elements in Eq.~(\ref{eq:sigma3L}) have argument $(-\omega_\Sigma;\omega_1,\omega_2,\omega_3)$. Eq.~(\ref{eq:sigma3L}) reduces to $\sigma^{(3)}_{\rm L}(-3\omega;\omega,\omega,\omega) =\sigma^{(3)}_{xxxx}(-3\omega;\omega,\omega,\omega)$ for the particular case of THG.

In an inversion symmetric system, we have 
$\sigma^{(2)}_{\ell \alpha_1\alpha_2}(-\omega_\Sigma;\omega_1,\omega_2) =0$.
However, a {\it non-vanishing} dipole $d^{(2)}_{\ell \alpha_1\alpha_2\beta}$ of the second-order conductivity is expected. Since $d^{(2)}_{\ell \alpha_1\alpha_2\beta}$ is a rank-$4$ tensor, it obeys the same symmetry properties of the third-order nonlinear conductivity. 

Because of the intrinsic permutation symmetry of the second-order conductivity tensor, we have
$
d^{(2)}_{\ell \alpha_1\alpha_2\beta,2}(-\omega_\Sigma;\omega_1,\omega_2)
=d^{(2)}_{\ell \alpha_2 \alpha_1\beta,1}(-\omega_\Sigma;\omega_2,\omega_1)
$.
Moreover, as demonstrated in Appendix~\ref{app:d2L}, $d^{(2)}_{xyyx,1}=d^{(2)}_{xxyy,1}$. 

We can therefore write the following result for $d^{(2)}_{\rm L,1}$:
\begin{eqnarray}\label{eq:d2L1}
&&d^{(2)}_{\rm L,1}(-\omega_\Sigma;\omega_1,\omega_2) =
d^{(2)}_{xyyx,1}  \cos(2\theta_1-\theta_2-\theta_\Sigma) 
\nonumber\\
&&+ \left [d^{(2)}_{xyxy,1} +d^{(2)}_{xyyx,1}  \right ]  \cos(\theta_\Sigma-\theta_2) ~.
\end{eqnarray}
All tensor elements on the right-hand side of Eq.~(\ref{eq:d2L1}) have argument $(-\omega_\Sigma;\omega_1,\omega_2)$. 
For the case of $d^{(2)}_{\rm L,2}$, we have (see Appendix \ref{app:d2L})
\begin{eqnarray}\label{eq:d2L2}
&&d^{(2)}_{\rm L,2} (-\omega_\Sigma;\omega_1,\omega_2) = 
 d^{(2)}_{xyyx,1}  \cos(2\theta_2-\theta_1-\theta_\Sigma) 
\nonumber\\
&&+\left(  d^{(2)}_{xyxy,1} +  d^{(2)}_{xyyx,1} \right )\cos(\theta_\Sigma-\theta_1) ~.
\end{eqnarray}

From now on, we use the symbol $d^{(2)}_{\ell \alpha_1\alpha_2\beta}$ as a shorthand for $d^{(2)}_{\ell \alpha_1\alpha_2\beta,1}$.

\section{Second-order optical conductivity and its dipole in 2D Dirac materials}
\label{sect:second-order-Dirac}

We are now ready to specialize our general results to the case of a specific material. 

We will consider a 2D Dirac material with two valleys, like graphene. The low-energy Hamiltonian reads as following~\cite{Katsnelson,kotov_rmp_2012}: ${\cal H}_\tau({\bm k})=\hbar v_{\rm F} (\tau k_x\sigma_x+k_y\sigma_y)$, where $v_{\rm F} \sim 10^{6}~{\rm m}/{\rm s}$ is the Fermi velocity, $\tau=\pm$ stands for the valley ($K$, $K^\prime$) index, and $\sigma_{x,y}$ represent ordinary Pauli matrixes acting in sublattice space. The eigenstates (i.e.~bands) of the this Hamiltonian are given $E^{\lambda}_{\bm k}=\lambda \hbar v_{\rm F} |{\bm k}|$, where $\lambda=\pm$ indicates conduction and valence bands. The corresponding eigenvectors are
$|\lambda, {\bm k}, \tau \rangle^{\rm T} \equiv [u^{\lambda}_\tau({\bm k})]^{\rm T} =\left [ 1 , \lambda \tau e^{i \tau \phi({\bm k})} \right ]/\sqrt{2}$,
where $\phi({\bm k})$ is the polar angle of the vector ${\bm k}$. The wavefunctions in real space are $\psi^\lambda_{{\bm k}\tau}({\bm r}) = u^{\lambda}_\tau({\bm k}) e^{i{\bm k}\cdot {\bm r}}/ \sqrt{{\cal S}}$ where ${\cal S}$ is the 2D electron system area. The matrix elements of the charge density ($\hat n$) and charge current ($\hat j_\alpha$) operators are given by
$
\big \langle \lambda', {\bm k}',\tau \big | \hat n({\bm q})  \big | \lambda, {\bm k},\tau \big \rangle                 \equiv
\big \langle \lambda', {\bm k}',\tau \big | e^{i{\bm q}\cdot {\bm r}} \big | \lambda, {\bm k},\tau \big \rangle 
$
and
$
\big \langle \lambda', {\bm k}',\tau \big | \hat j_{\alpha}({\bm q}) \big | \lambda, {\bm k}, \tau \big \rangle  \equiv
\frac{1}{2} 
\big \langle \lambda', {\bm k}',\tau \big |  \big \{ j_{\alpha}, e^{i{\bm q}\cdot {\bm r}} \big\} \big | \lambda, {\bm k},\tau \big \rangle
$
where $\{\dots ,\dots\}$ stands for the anti-commutation and the paramagnetic current operator in the first quantization picture reads
$
j_\alpha = -( {e}/{\hbar} ) {\partial {\cal H}_{\tau}({\bm k})}/{\partial k_\alpha}~.
$
Therefore, $(j_x , j_y) =-e v_{\rm F} (\tau \sigma_x , \sigma_y)$. 

In the scalar potential gauge, the second-order charge current reads as follows~\cite{Habib_prb_2016}:
\begin{eqnarray}
J^{(2)}_{\ell}({\bm q_\Sigma},\omega_\Sigma) &=& \Pi^{(2)}_\ell(-{\bm q}_\Sigma,{\bm q}_1,{\bm q}_2,-\omega_\Sigma,\omega_1,\omega_2) 
\nonumber \\ &\times&
V({\bm q}_1,\omega_1) V({\bm q}_2,\omega_2)~.
\end{eqnarray}
Here, $V({\bm q},\omega)$ denotes the Fourier transform of the external scalar potential and $\Pi^{(2)}_\ell$ is the second-order response function that establishes a link between the current response $J^{(2)}_{\ell}({\bm q_\Sigma},\omega_\Sigma)$ and the product of two external scalar potentials, $V({\bm q}_1,\omega_1)$ and $V({\bm q}_2,\omega_2)$. The quantity $\Pi^{(2)}_\ell$ is diagrammatically represented by a triangular diagram similar to the one in Fig.~\ref{fig:chi2_RPA}a), with two density vertices and one current vertex, i.e.
\begin{eqnarray}\label{eq:pi2}
&&\Pi^{(2)}_\ell(-{\bm q}_\Sigma,{\bm q}_1,{\bm q}_2,-\omega_\Sigma,\omega_1,\omega_2) = 
  \frac{-e^3 v_{\rm F}}{\cal S}\sum_{\bm k} \sum_{\{\lambda_i\}}  \sum'_{\cal P}
    \nonumber\\ &&
  \frac{ 
   F_{\ell, \lambda_1\lambda_2\lambda_3}({\bm k},{\bm q}_1,{\bm q}_2)}{\hbar\omega_\Sigma+ E_{\lambda_1, {\bm k} }-E_{\lambda_3,{\bm k} +{\bm q}_\Sigma}} 
   \Bigg [ 
    \frac{n_{\rm F} \left (E_{\lambda_1, {\bm k}} \right )-n_{\rm F} \left (E_{\lambda_2,{\bm k} +{\bm q}_1} \right ) }{ {\hbar\omega_1+E_{\lambda_1, {\bm k}}-E_{\lambda_2,{\bm k} +{\bm q}_1}  }} 
    - \nonumber\\ &&
    \frac{n_{\rm F} \left (E_{\lambda_2, {\bm k} +{\bm q}_1} \right )-n_{\rm F} \left (E_{\lambda_3,{\bm k} +{\bm q}_\Sigma} \right ) }{ {\hbar\omega_2+E_{\lambda_2, {\bm k} +{\bm q}_1}-E_{\lambda_3,{\bm k} +{\bm q}_\Sigma}}}  
   \Bigg ]~.
\end{eqnarray}
In Eq.~(\ref{eq:pi2})  we have introduced the form factor
\begin{eqnarray}\label{eq:form-factor}
F_{x, \lambda_1\lambda_2\lambda_3}({\bm k},{\bm q}_1,{\bm q}_2) &=&
\frac{\lambda_1 e^{-i\tau\phi({\bm k})} +\lambda_3  e^{i\tau\phi({\bm k}+{\bm q}_\Sigma)}}{2}
\nonumber\\
&\times&\frac{1+\lambda_2 \lambda_3 e^{-i\tau[\phi({\bm k}+{\bm q}_\Sigma)-\phi({\bm k}+{\bm q}_1)]}}{2}
\nonumber\\
&\times&\frac{1+\lambda_1 \lambda_2 e^{-i\tau[\phi({\bm k}+{\bm q}_1)-\phi({\bm k})]}}{2}~.
\end{eqnarray}
As customary in many-body perturbation theory~\cite{Giuliani_and_Vignale}, everywhere in Eq.~(\ref{eq:pi2}) $\hbar\omega_i$ denotes a photon energy accompanied by an infinitesimal positive imaginary part, i.e.~$\hbar\omega_i \to \hbar\omega_i +i \eta$, with $\eta=0^{+}$.
The symbol $\sum'_{\cal P}$ implies that we are enforcing the intrinsic permutation symmetry\cite{Butcher_and_Cotter} between $({\bm q}_1,\omega_1)$ and $({\bm q}_2,\omega_2)$.

Because of inversion symmetry, the second-order optical conductivity $\sigma^{(2)}_{\ell \alpha_1\alpha_2}(-\omega_\Sigma;\omega_1,\omega_2)$ is identically zero in the case of graphene. (It is non-zero in the case of a finite valley polarization and can be used as a diagnostic tool for the presence of the latter~\cite{wehling_prb_2015}.) However, its dipole is finite:
\begin{equation}\label{eq:d2_derv}
d^{(2)}_{\ell \alpha_1\alpha_2\beta}  = i^2 \frac{\partial^3 \Pi^{(2)}_{\ell}}{\partial q_{1,\alpha_1} \partial q_{2,\alpha_2} \partial q_{1,\beta}} \Bigg |_{ \{{\bm q}_1,{\bm q}_2\} \to 0}\neq 0~.
\end{equation}
After straightforward algebraic steps (which are summarized in Appendix~\ref{app:d2xxxx}), we obtain the following expression for $d^{(2)}_{xxxx}$ at zero temperature:
\begin{eqnarray}\label{eq:d2xxxx}
d^{(2)}_{xxxx}(-\omega_\Sigma;\omega_1,\omega_2) &=&  \frac{d_0}{\hbar\omega_1\hbar\omega_2} 
 \left ( \frac{1}{\hbar\omega_1}+\frac{1}{\hbar\omega_\Sigma} \right ) 
 \\ & \times&
  \frac{16 E_{\rm F}^4 }{\left [(\hbar\omega_1)^2-4 E^2_{\rm F} \right ]  \left [ (\hbar\omega_\Sigma)^2 -4 E^2_{\rm F} \right ]} 
  \nonumber
\end{eqnarray}
where we have introduced
\begin{equation}\label{eq:dzero}
d_0 \equiv {\rm sign}(E_{\rm F})\frac{e^3\hbar v^2_{\rm F}}{4\pi}~.
\end{equation}
We have checked (not shown here) that the expected permutation symmetry $d^{(2)}_{xxxx,2}(-\omega_\Sigma;\omega_1,\omega_2) =d^{(2)}_{xxxx,1}(-\omega_\Sigma;\omega_2,\omega_1)$ is satisfied. Our results in Eqs.~(\ref{eq:d2xxxx})-(\ref{eq:dzero}) coincide with those recently reported in Ref.~\onlinecite{Tokman_prb_2016}.

The other non-vanishing tensor elements of the second-order dipole are: 
\begin{eqnarray}
&&d^{(2)}_{xyyx}=d^{(2)}_{xxyy} =  
\frac{{d_0}/{(\hbar\omega_1\hbar\omega_2)}}{\hbar \omega_1\hbar \omega_\Sigma\left [(\hbar\omega_1)^2-4 E^2_{\rm F} \right ]  \left [ (\hbar\omega_\Sigma)^2 -4 E^2_{\rm F} \right ]}
\nonumber\\ &&\times
8 E^2_{\rm F} \left [  (\hbar\omega_1)^2 \hbar\omega_\Sigma -2\hbar\omega_2 E^2_{\rm F}\right ]
\end{eqnarray}
and
\begin{eqnarray}
&&d^{(2)}_{xyxy} =  
\frac{{d_0}/{(\hbar\omega_1\hbar\omega_2)}}{\hbar \omega_1\hbar \omega_\Sigma\left [(\hbar\omega_1)^2-4 E^2_{\rm F} \right ]  \left [ (\hbar\omega_\Sigma)^2 -4 E^2_{\rm F} \right ]}
\nonumber \\ && \times
16 E^2_{\rm F} \left [ E^2_{\rm F} (2\hbar\omega_1+3\hbar\omega_2)-(\hbar\omega_1)^2 \hbar\omega_\Sigma\right ]
~.
\end{eqnarray}
It can be shown that the identity $d^{(2)}_{xxxx}=d^{(2)}_{xyxy}+d^{(2)}_{xyyx}+d^{(2)}_{xxyy}$ holds true.

\begin{figure}[t]
\centering
\includegraphics[width=0.9\linewidth]{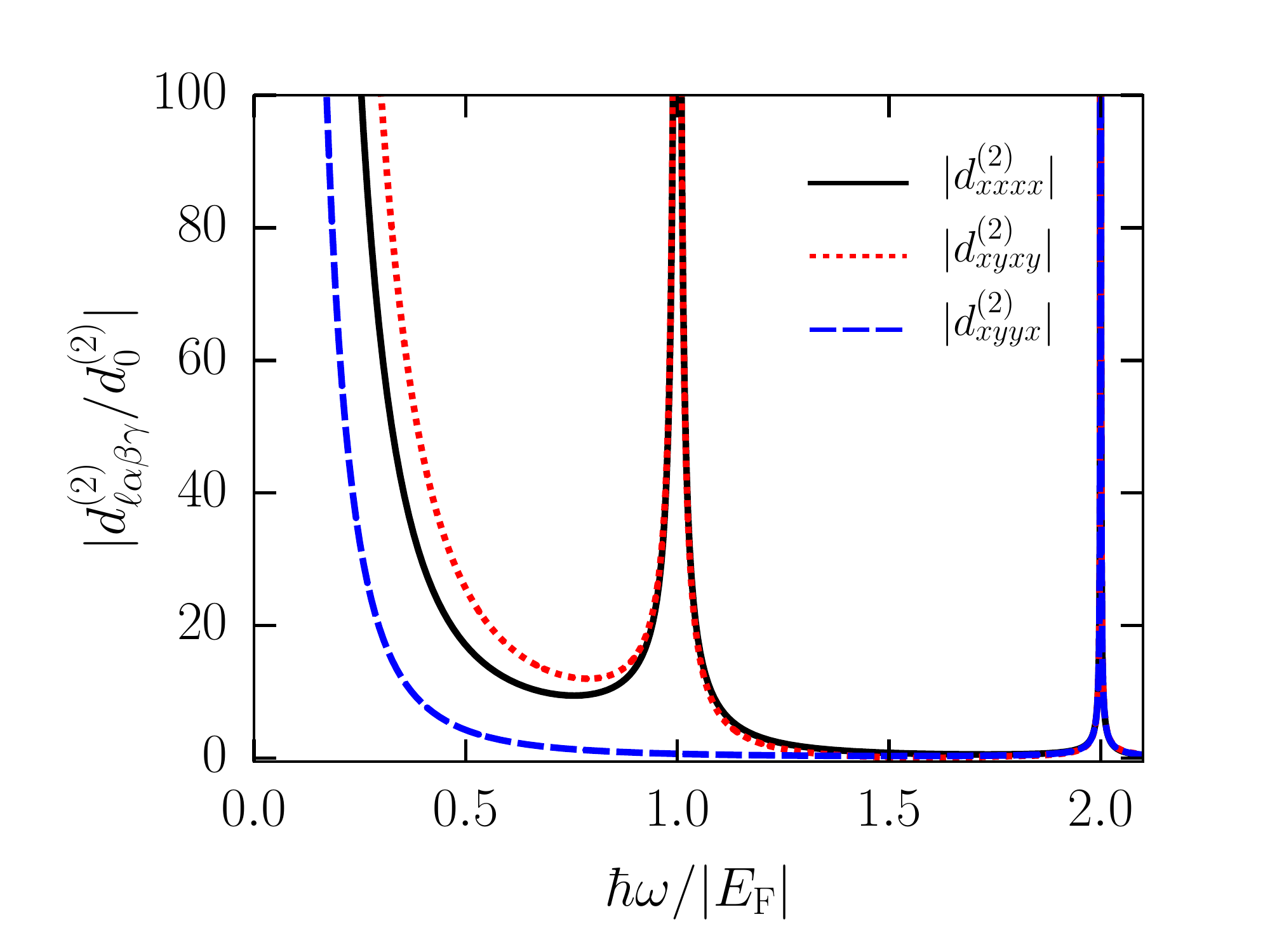} 
\caption{(Color online) Illustrative plots of $|d^{(2)}_{xxxx}(-2\omega;\omega,\omega)|$, $|d^{(2)}_{xyxy}(-2\omega;\omega,\omega)|$, and $|d^{(2)}_{xyyx}(-2\omega;\omega,\omega)|$---in units of $d^{(2)}_0= d_0/|E_{\rm F}|^3$, where $d_{0}$ has been introduced in Eq.~(\ref{eq:dzero})---for the case of clean graphene, at zero temperature. The quantities $d^{(2)}_{xxxx}$ and $d^{(2)}_{xyxy}$ (solid black and dotted red lines) show {\it two} sharp resonances at $\hbar \omega=|E_{\rm F}|$ and $\hbar \omega = 2|E_{\rm F}|$, while $d^{(2)}_{xyyx}$ (dashed blue line) shows only one resonance at $\hbar\omega=2 |E_{\rm F}|$.\label{fig:d2}}
\end{figure}
\section{Plasmon-dressed second- and third-harmonic generation in 2D Dirac materials}
\label{sect:harmonic}
\begin{figure}[t]
\centering
 \begin{overpic}[width=0.81\linewidth]{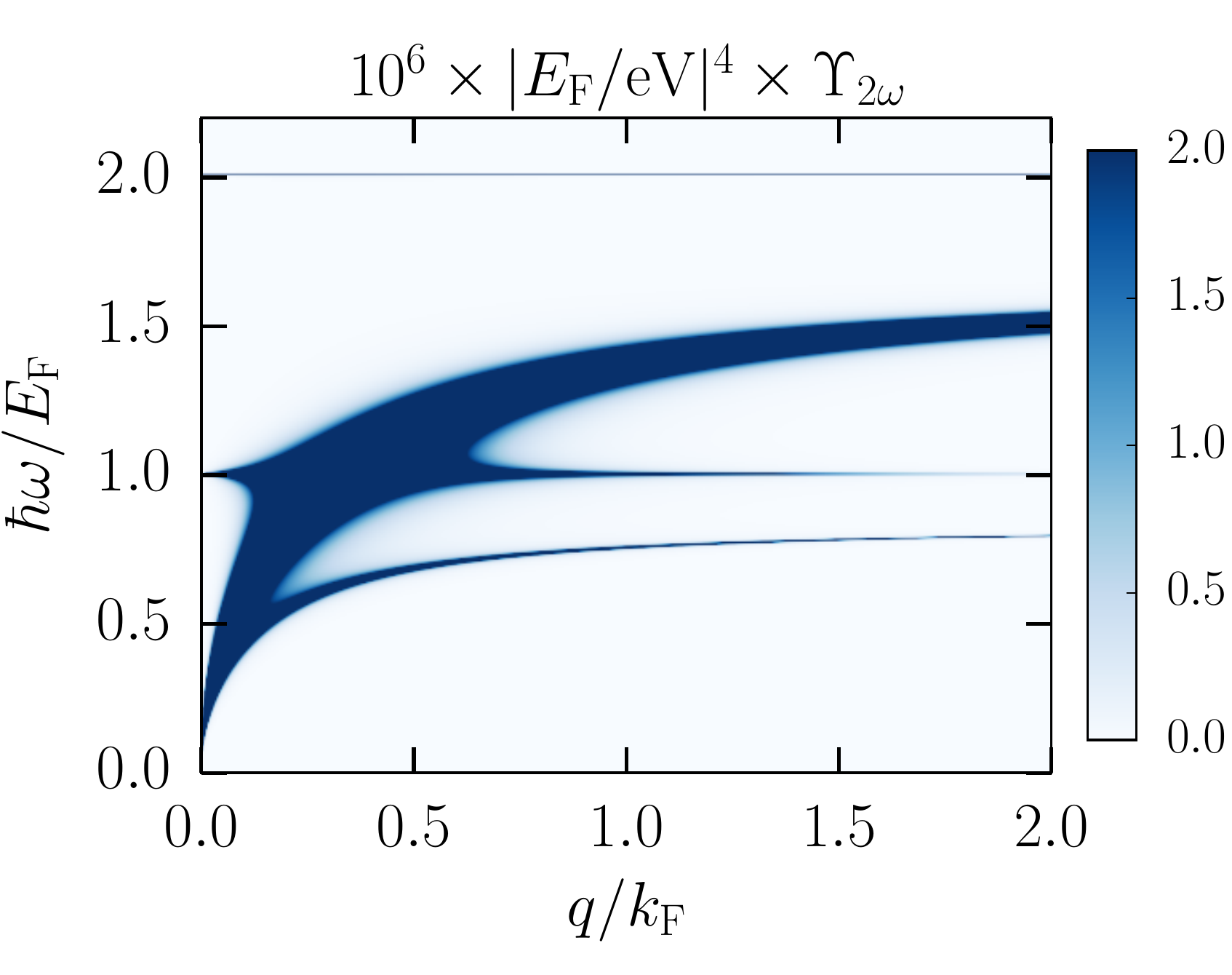}\put(-1,67){(a)}\end{overpic}
 \begin{overpic}[width=0.81\linewidth]{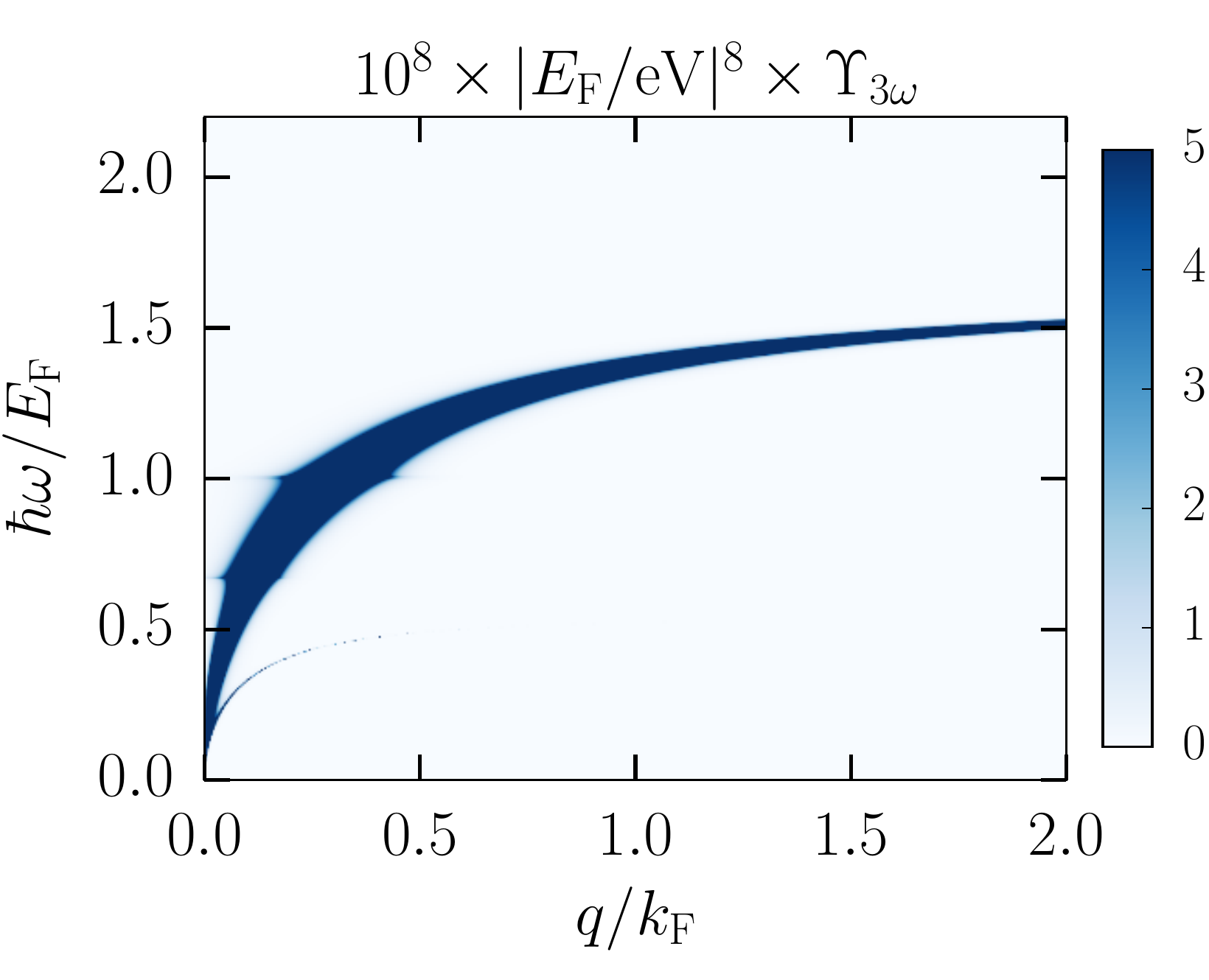}\put(-1,67){(b)}\end{overpic}
 \begin{overpic}[width=0.81\linewidth]{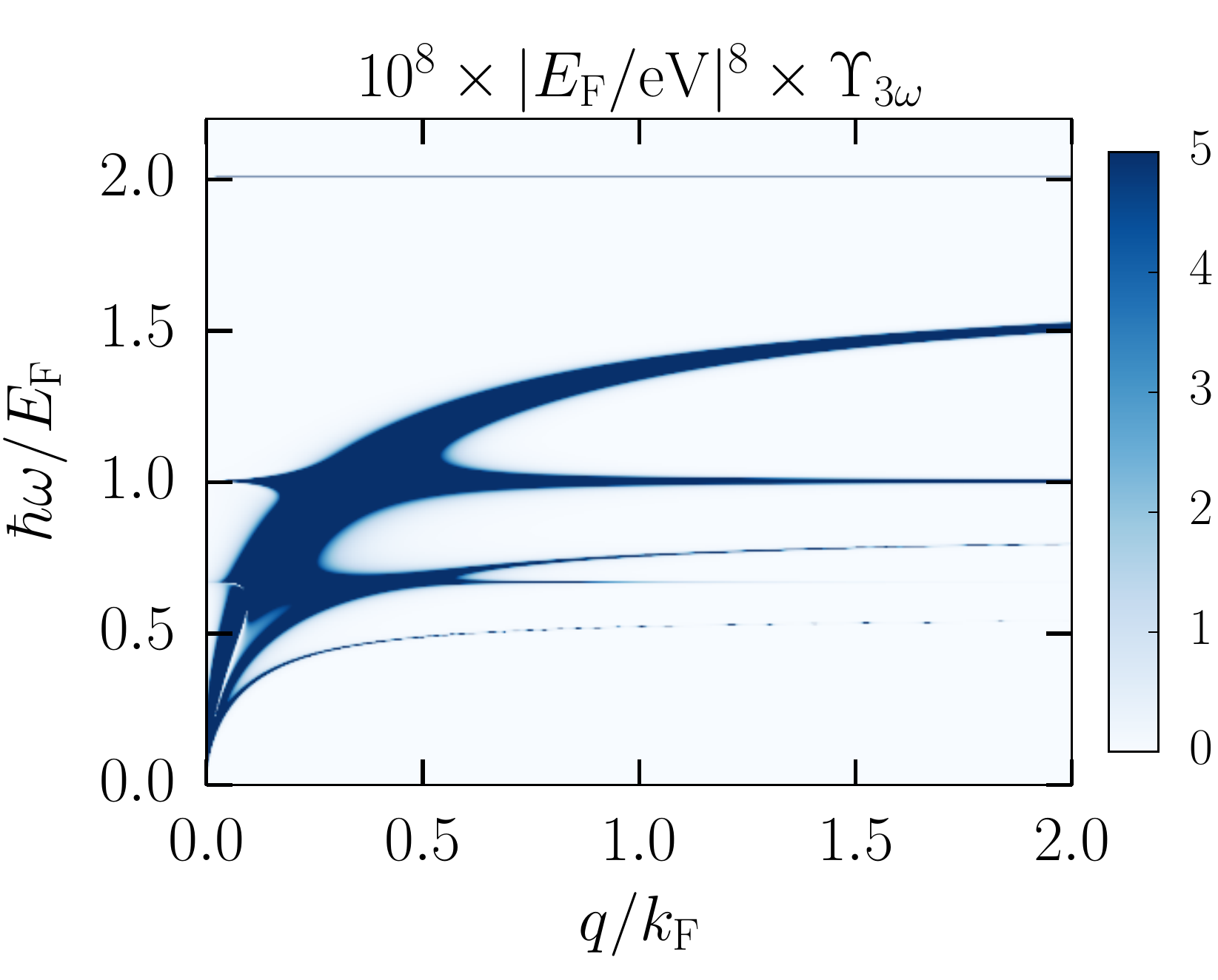}\put(-1,67){(c)}\end{overpic}
 \caption{(Color online) Dimensionless efficiencies for harmonic generation in graphene are shown as functions of the wavevector $q$ (in units of $k_{\rm F}$) and $\omega$ (in units of $E_{\rm F}/\hbar$). These plots have been made by taking $I_{\rm in}=1~{\rm GW/cm}^2$. Panel (a) illustrates the SHG efficiency $\Upsilon_{2\omega}$, as dressed by plasmons, i.e.~Eq.~(\ref{eq:Ishg-Iin}). Panel (b) illustrates the THG efficiency $\Upsilon_{3\omega}$, in the case in which one sets ${\widetilde \sigma}^{(3)}_{xxxx} = 0$ in Eq.~(\ref{eq:plasmon-THG}). Panel (c) illustrates the same quantity as in panel (b), but for ${\widetilde \sigma}^{(3)}_{xxxx} \neq 0$.\label{fig:plasmonic_shg_thg}}
\end{figure}

We first consider the case of a {\it single} laser beam with frequency $\omega$ and wavevector ${\bm q}$, which is polarized along the $\hat{\bm x}$ direction. 

According to Eq.~(\ref{eq:sigma2_eff}), the nonlinear conductivity for the case of second-harmonic generation (SHG) is 
\begin{equation}\label{eq:plasmon-SHG}
\sigma^{{\rm SHG}, {\rm ee}}  = 2 q \frac{d^{(2)}_{xxxx}(-2\omega;\omega,\omega)}{{\cal R}_{2}} +\dots~,
\end{equation}
where ${\cal R}_2$ has been introduced in Eq.~(\ref{eq:R_n}) and encodes the plasmonic enhancement of the second-order response due to the collective behavior of the many-electron system. Notice that $\sigma^{{\rm SHG}, {\rm ee}}$ is manifestly zero in the homogeneous $q=0$ limit, in agreement with the fact that graphene is an inversion-symmetric material.

Illustrative plots of the second-order conductivity dipole in the case of SHG are reported in Fig.~\ref{fig:d2}. We clearly see that $d^{(2)}_{xxxx}$ and $d^{(2)}_{xyxy}$ display {\it two} sharp resonances at $\hbar \omega = |E_{\rm F}|$ and $\hbar \omega = 2|E_{\rm F}|$, while $d^{(2)}_{xyyx}$ displays only {\it one} resonance at $\hbar\omega=2 |E_{\rm F}|$. 

The calculation of ${\cal R}_2$ requires knowledge of the first-order optical conductivity, which in the case of clean graphene at zero temperature reads as following~\cite{koppens_nanolett_2011,basov_rmp_2014}:
\begin{equation}
\sigma^{(1)}_{xx} (\omega)  =  i \sigma_{\rm uni} 
\left \{
 \frac{4|E_{\rm F}|}{ \pi \hbar \omega} 
-\frac{1}{\pi} \ln \left [ \frac{2 |E_{\rm F}|+\hbar\omega}{2 |E_{\rm F}|-\hbar\omega}\right ]
\right \}~.
\end{equation}
Here, $\sigma_{\rm uni} = e^2/(4\hbar)$ is the universal optical conductivity~\cite{koppens_nanolett_2011,grigorenko_naturephoton_2012,basov_rmp_2014}.

We now estimate the dimensionless {\it efficiency} $\Upsilon_{2\omega}$ of second-harmonic nonlinear processes, as dressed by electron-electron interaction effects. 
Using a slowly-varying envelope approximation approach and neglecting the so-called ``wavevector mismatch''~\cite{Butcher_and_Cotter}, we find 
\begin{equation}\label{eq:Ishg-Iin}
\Upsilon_{2\omega}\equiv \frac{I_{2\omega}}{I_{\rm in}} \simeq \frac { I_{\rm in} }{8 n^2_{\omega} n_{2\omega} \epsilon^3_0 c^3} \left |\sigma^{{\rm SHG}, {\rm ee}}\right |^2~,
\end{equation}
where $I_{\rm in}$ and $I_{2\omega}$ stand for incident-beam and second-harmonic-generated signal intensities, respectively. Notice that $n_{\omega} \approx n_{2\omega} \approx 1$ is the real part of graphene refraction index. Illustrative plots of $\Upsilon_{2\omega}$ for the case of clean graphene at zero temperature are reported in Fig.~\ref{fig:plasmonic_shg_thg}(a). We clearly see two plasmon-related singularities, one located at the usual Dirac plasmon pole, i.e.~at $\omega = \omega_{\rm p}(q) \propto \sqrt{q}$, and one at 
$\omega = \omega_{\rm p}(q)/\sqrt{2}$. We note that $\omega_{\rm p}(q)/\sqrt{n}$ is the 
root of the equation ${\rm Re}[1 + i  q / (2\epsilon_0 \omega) \sigma^{(1)}_{\rm L}(n \omega) ] =0$ for $n=1,2,3,{\rm ~etc}$. In the long-wavelength limit and in the case of a free-standing graphene sheet, linear-response theory in the RPA yields~\cite{grigorenko_naturephoton_2012} $\omega_{\rm p}(q \ll k_{\rm F}) = \sqrt{{\cal D}_{0} q/(\pi\epsilon_{0})}$, where ${\cal D}_{0} = 4\sigma_{\rm uni}|E_{\rm F}|/\hbar$ is the non-interacting Drude weight in graphene~\cite{grigorenko_naturephoton_2012}. We also note that the resonance at $\omega = \omega_{\rm p}(q)$ is much stronger than the one at $\omega = \omega_{\rm p}(q)/\sqrt{2}$. The reason is that the former is due to a second-order pole, i.e.~$1/{\cal R}_2 \propto [\omega-\omega_{\rm p}(q)]^{-2}$ near $\omega = \omega_{\rm p}(q)$, while the latter one is due to a first-order pole, i.e.~$1/{\cal R}_2 \propto [\omega-\omega_{\rm p}(q)/\sqrt{2}]^{-1}$ near $\omega = \omega_{\rm p}(q)/\sqrt{2}$. Of course, $\Upsilon_{2\omega}$ contains the same non-plasmonic poles of $d^{(2)}_{xxxx}(-2\omega,\omega,\omega)$, which are located at $\omega=|E_{\rm F}|/\hbar$ 
and $\omega= 2 |E_{\rm F}|/\hbar$.

We now proceed to discuss plasmon-related effects in the third-harmonic nonlinearity. 
Using Eq.~(\ref{eq:sigma3_eff}),  the dressed third-harmonic conductivity in the long-wavelength limit reads as following:
\begin{equation}\label{eq:plasmon-THG}
\sigma^{{\rm THG}, {\rm ee}}\equiv  \frac{\sigma^{\rm (3)}_{xxxx}+ \widetilde\sigma^{\rm (3)}_{xxxx}}{{\cal R}_3} +\dots~,
\end{equation}
where $ \widetilde\sigma^{\rm (3)}_{xxxx}$ is defined in 
Eq.~(\ref{eq:tilde_sigma3}) and ${\cal R}_{3}$ in Eq.~(\ref{eq:R_n}). 

For the case of harmonic generation and in the long-wavelength limit, we have
\begin{eqnarray} \label{eq:tilde_sigma3_xxxx}
{\widetilde \sigma}^{\rm (3)}_{xxxx}  &=&
-i \frac{3  q^3 }{\epsilon_0 \omega}\frac{d^{(2)}_{xxxx}(-2\omega;\omega,\omega)}
{1 + i q \sigma^{(1)}_{xx}(2\omega)/(2\epsilon_{0} \omega)} 
\Big [ d^{(2)}_{xxxx}(-3\omega;\omega,2\omega) 
\nonumber \\&+&2 d^{(2)}_{xxxx}(-3\omega;2\omega,\omega) 
\Big ] +\dots 
\end{eqnarray}
In the case of graphene, $\sigma^{\rm (3)}_{xxxx}$ displays~\cite{Habib_prb_2016,cheng_prb_2015,mikhailov_prb_2016} three weak (i.e.~{\it logarithmic}) singularities at $\hbar\omega = 2|E_{\rm F}|/3$, $|E_{\rm F}|$, and $2|E_{\rm F}|$.  On the other hand, the second-order dipole contributions $d^{(2)}_{xxxx}(-2\omega;\omega,\omega)$, $d^{(2)}_{xxxx}(-3\omega;\omega,2\omega)$, and $d^{(2)}_{xxxx}(-3\omega;2\omega,\omega)$ are responsible for much stronger singularities (i.e.~first-order poles) in ${\widetilde \sigma}^{\rm (3)}_{xxxx}$, at the same resonant frequencies. This implies that the electron-electron interaction contribution ${\widetilde \sigma}^{\rm (3)}_{xxxx}$ to the third-order conductivity can be much larger than the bare contribution $\sigma^{\rm (3)}_{xxxx}$ when $\omega$ approaches the resonant frequencies, despite the small (but finite) $q^3$ factor in Eq.~(\ref{eq:tilde_sigma3_xxxx}).  Leaving aside these resonances of single-particle origin, the dressed THG conductivity $\sigma^{{\rm THG}, {\rm ee}}$ in Eq.~(\ref{eq:plasmon-THG}) displays plasmon-related poles. These are due to the explicit factor $1/{\cal R}_{3}$ in Eq.~(\ref{eq:plasmon-THG}), but also due to the denominator $ 1 + i   q\sigma^{(1)}_{xx}(2\omega)/(2\epsilon_{0} \omega)$ in Eq.~(\ref{eq:tilde_sigma3_xxxx}). This latter factor, in particular, is also responsible for a further enhancement of ${\widetilde \sigma}^{\rm (3)}_{xxxx}$ with respect to the bare value $\sigma^{\rm (3)}_{xxxx}$.

Following the same steps that led to Eq.~(\ref{eq:Ishg-Iin}), we reach the following estimate for the third-harmonic conversion efficiency:
\begin{equation}\label{eq:Ithg-Iin}
\Upsilon_{3\omega} \equiv \frac{I_{3\omega}}{I_{\rm in}} \approx \frac { I^2_{\rm in} }{16 n^3_{\omega} n_{3\omega} \epsilon^4_0 c^4} \left |\sigma^{{\rm THG}, {\rm ee}} \right |^2~.
\end{equation}
Illustrative plots of $\Upsilon_{3\omega}$ for the case of clean graphene at zero temperature are reported in Figs.~\ref{fig:plasmonic_shg_thg}(b) and~\ref{fig:plasmonic_shg_thg}(c). In particular, in obtaining the results shown in Fig.~\ref{fig:plasmonic_shg_thg}(b), we have deliberately set ${\widetilde \sigma}^{(3)}_{xxxx} = 0$ in Eq.~(\ref{eq:plasmon-THG}), for illustrative purposes only. Results for the case ${\widetilde \sigma}^{(3)}_{xxxx} \neq 0$ are presented in Fig.~\ref{fig:plasmonic_shg_thg}(c). 

In Fig.~\ref{fig:plasmonic_shg_thg}(b), we clearly see that the quantity $\Upsilon_{3\omega}$ is large at $\omega = \omega_{\rm p}(q)$ and that it displays a weaker plasmon ``satellite'' at $\omega = \omega_{\rm p}(q)/\sqrt{3}$. We can further notice two more poles at $\hbar \omega=2|E_{\rm F}|/3$ and $\hbar\omega=|E_{\rm F}|$, which are visible only when they merge with the main plasmon branch and are otherwise too weak to be seen. These poles originate from the aforementioned logarithmic singularities~\cite{Habib_prb_2016,cheng_prb_2015,mikhailov_prb_2016} of the bare THG conductivity 
$\sigma^{(3)}_{xxxx}$. 

In Fig.~\ref{fig:plasmonic_shg_thg}(c), we include also the effect of $\widetilde\sigma^{(3)}_{xxxx}$ on $\Upsilon_{3\omega}$. Considering this interaction-induced nonlocal modification of the third-harmonic conductivity, we find a dramatic enhancement of all plasmonic and single-particle poles of the THG efficiency. We also see that, because of $\widetilde\sigma^{(3)}_{xxxx}$, an extra plasmon satellite appears at $\omega = \omega_{\rm p}(q)/\sqrt{2}$. The latter emerges from the denominator in Eq.~(\ref{eq:tilde_sigma3_xxxx}). Moreover, after taking $\widetilde\sigma^{(3)}_{xxxx}$ into account, a single-particle pole at $\hbar\omega=2|E_{\rm F}|$ shows up, while the peaks that were barely visible in Fig.~\ref{fig:plasmonic_shg_thg}(b) at $\hbar \omega = 2|E_{\rm F}|/3 $ and $|E_{\rm F}|$ become much stronger. 
\section{Plasmon-dressed sum- and difference-frequency wave mixing in 2D Dirac materials}
\label{sect:sum_and_difference}
We now turn to consider an experimental setup with {\it two} laser beams with frequencies $\omega_{1}$ and $\omega_{2}$. 
The wavevector of each beam has in-plane, ${\bm q}_i$, and perpendicular-to-the-plane, ${\bm q}^{\perp}_i$, components (with respect to the plane of the 2D electron system). 
Notice that $q_i=|{\bm q}_i|=\omega_i \cos(\vartheta_i)/c$ and  $q^{\perp}_i=|{\bm q}^{\perp}_i|=\omega_i \sin(\vartheta_i)/c$ where $c$ is the speed of light and $\vartheta_{i}$ is the angle between the $i$-th beam and the plane where graphene lies.
\par
\begin{figure}[t]
\centering
\begin{overpic}[width=0.905\linewidth]{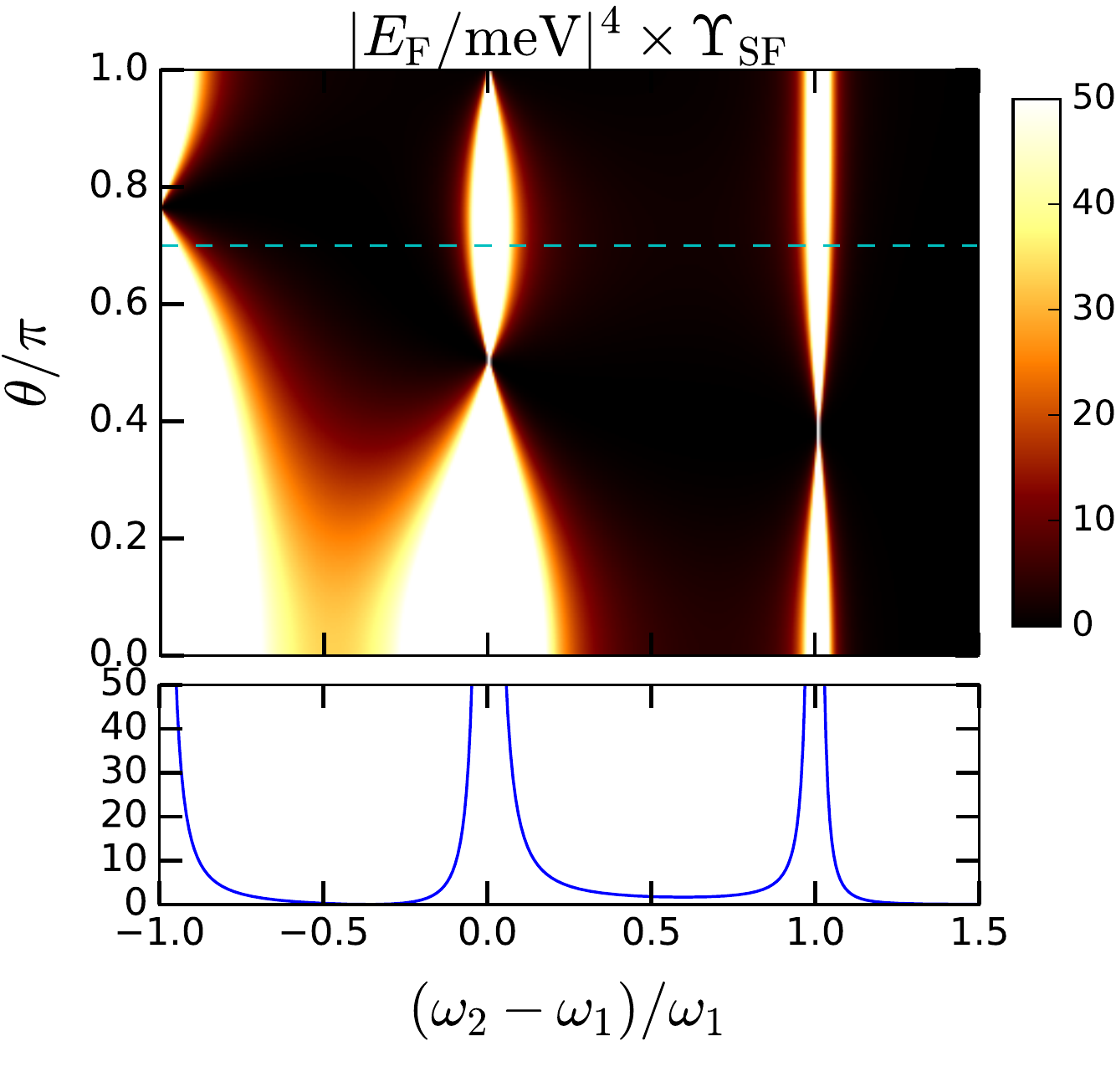}\put(-1,90){(a)}\end{overpic}
\begin{overpic}[width=0.91\linewidth]{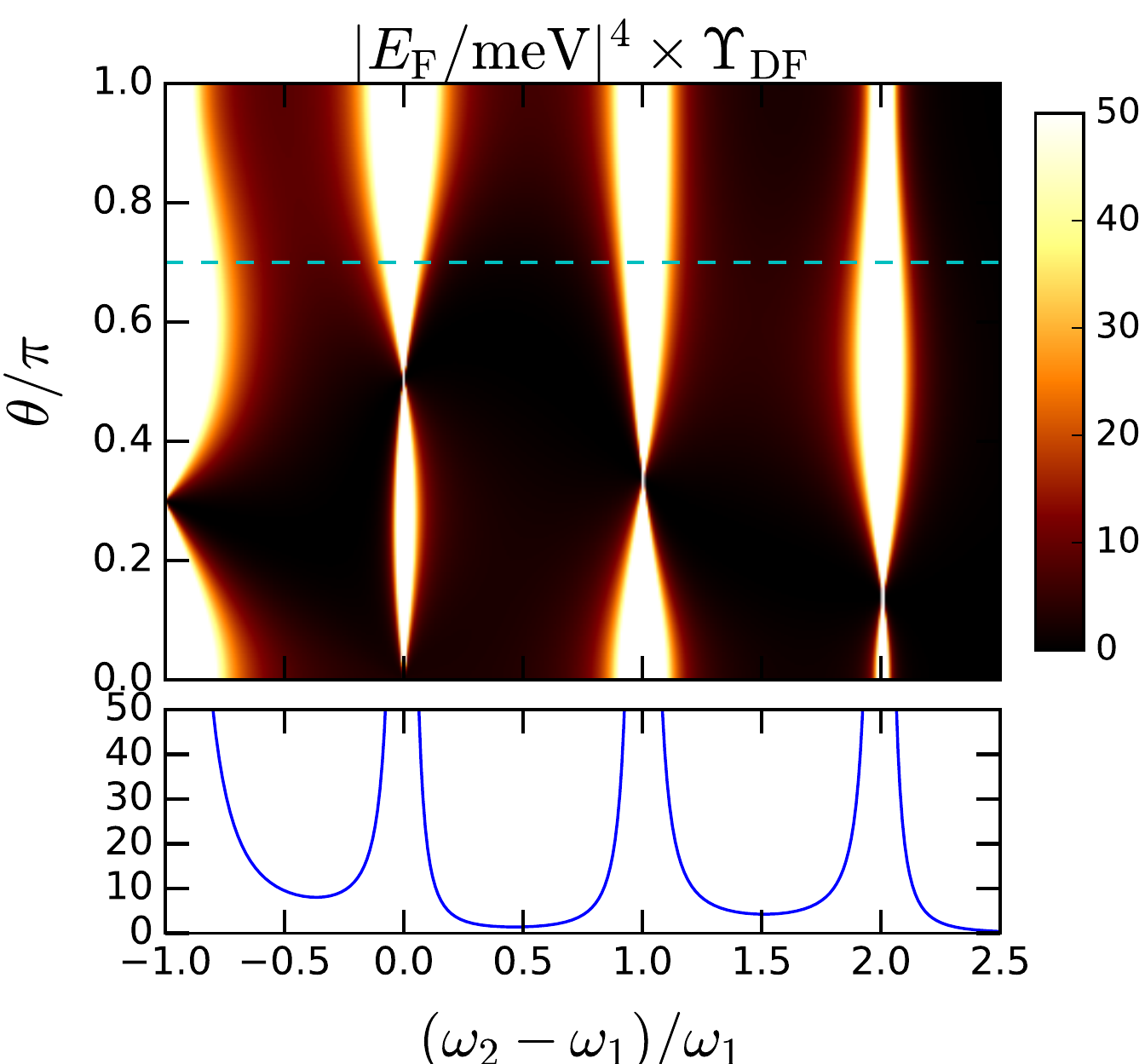}\put(-1,85){(b)}\end{overpic}
\caption{(Color online) Color maps of the {\it bare} sum-frequency and difference-frequency wave mixing conversion efficiencies in graphene. The quantities $\Upsilon_{\rm SF}$---panel (a)---and $\Upsilon_{\rm DF}$---panel (b)---are plotted as functions of the relative angle $\theta$ and relative frequency $(\omega_{2} - \omega_{1})/\omega_{1}$. In making this plot we have neglected the effect of electron-electron interactions by setting ${\cal R}_2=1$. Also, we have set $I_{1}=I_{2} = 1{\rm GW/ cm^2} $, $\hbar\omega_{1}=E_{\rm F}$, and $q_i=\omega_i/c$. For the sake of simplicity, we have neglected the perpendicular component of all wavevectors in the previous relation. In each panel, we have also reported a one-dimensional cut of the 2D color map, taken along the dashed green line at $\theta = 0.7\pi$.\label{fig:YSFDFNof2}}
\end{figure}
\begin{figure}[t]
\begin{overpic}[width=0.9\linewidth]{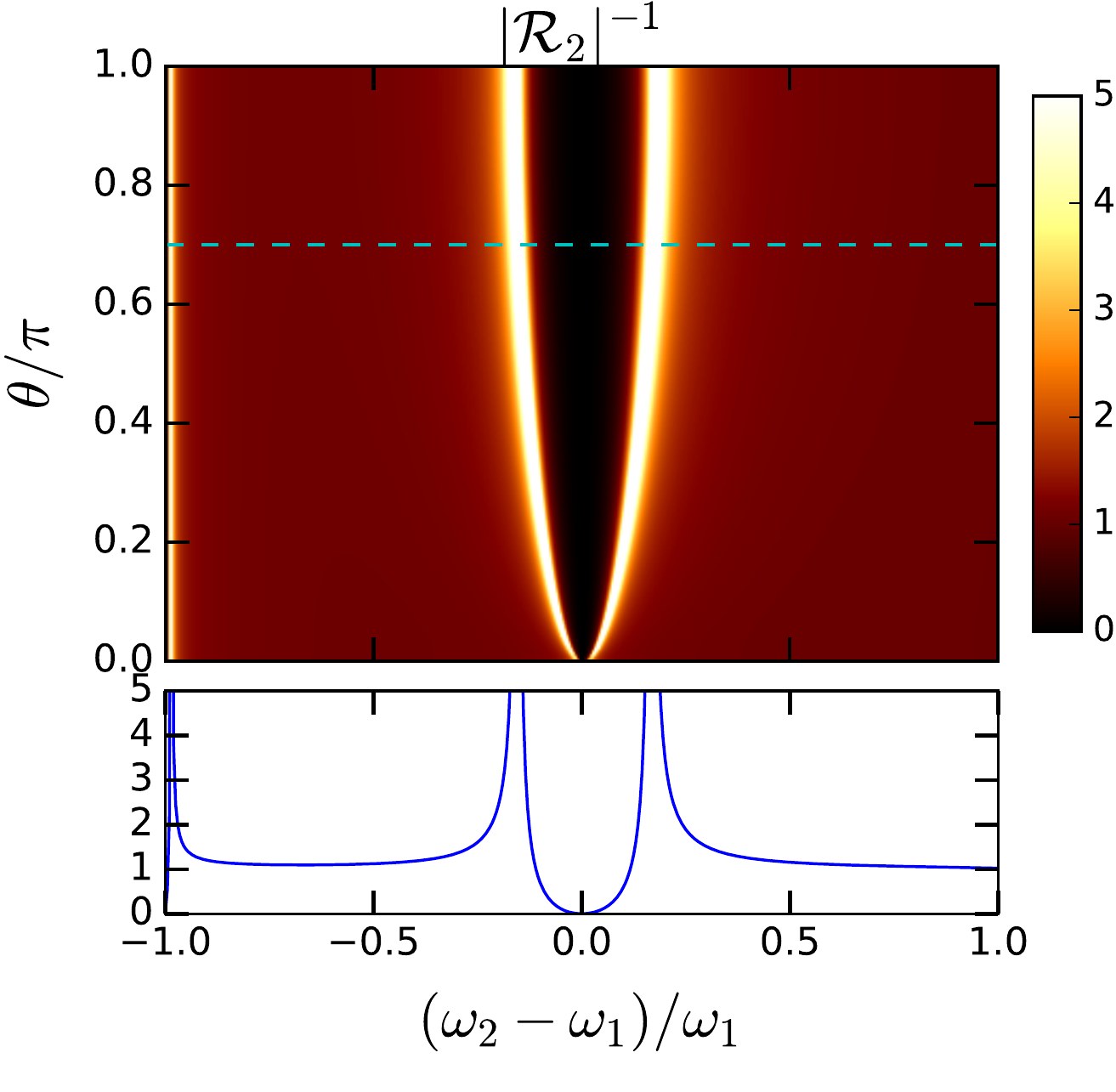}\put(-1,89.5){ (a)}\end{overpic}
\begin{overpic}[width=0.9\linewidth]{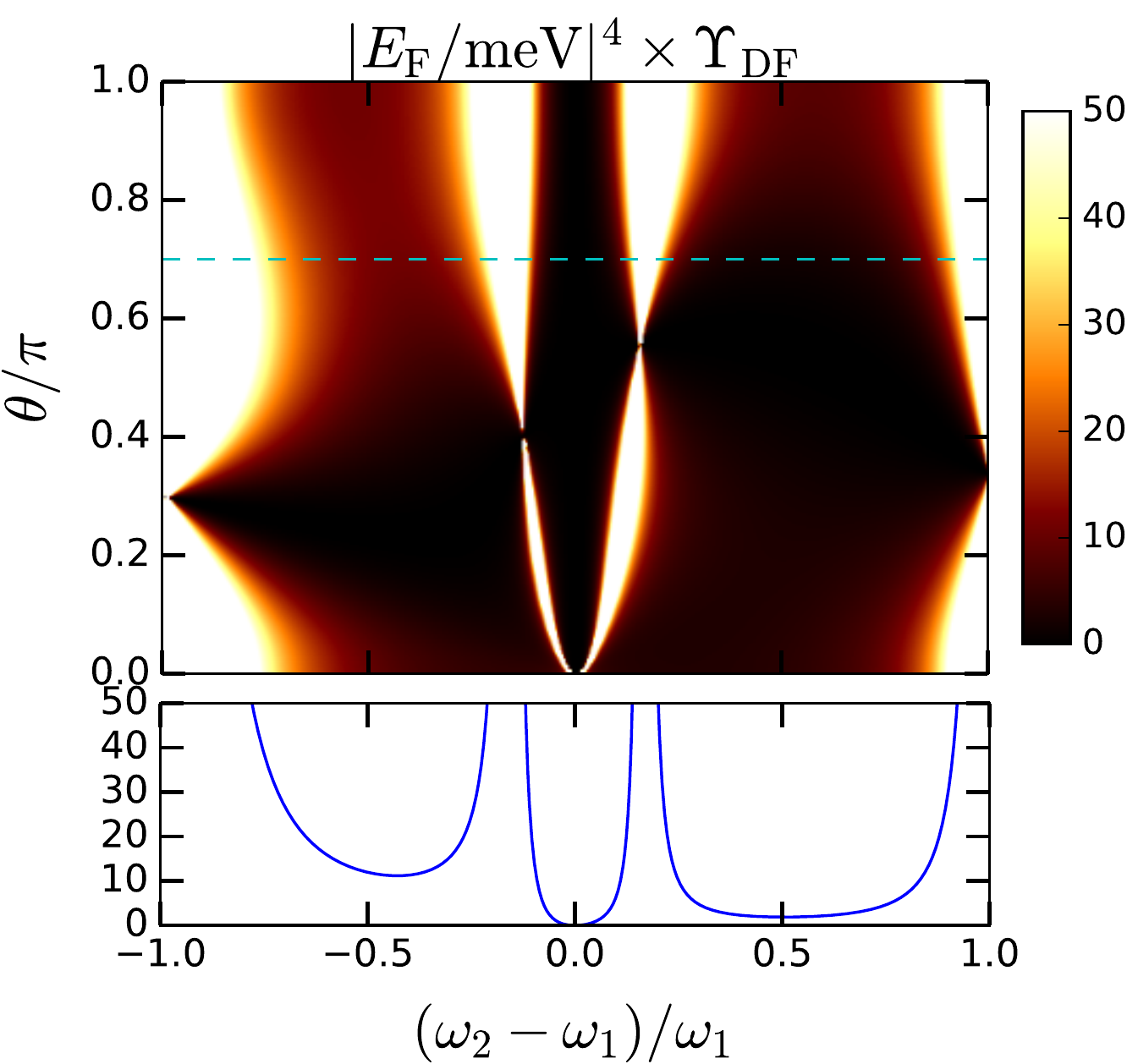}\put(-1,87){ (b)}\end{overpic}
\caption{(Color online)  Panel (a) Color map of the dimensionless quantity $|{\cal R}_2|^{-1}$ for the case of difference-frequency wave mixing. Here, $\hbar\omega_{1} = E_{\rm F}$ and $q_i=\omega_i/c$.  We clearly see two sharp plasmon resonances. Also, we notice that $|{\cal R}_2|^{-1}$ drops quickly to zero away from these resonances due to the low-frequency Drude peak in the first-order conductivity.  Panel (b) Difference-frequency wave mixing conversion efficiency with the inclusion of electron-electron interactions. The reader is invited to compare this panel with panel (b) in Fig.~\ref{fig:YSFDFNof2}. 
\label{fig:YSFDFplasmon}}
\end{figure}

Using again a slowly-varying envelope approximation approach~\cite{Butcher_and_Cotter}, one can show that the output electrical signal ${\cal E}^{\rm out}_\ell$ is proportional to $\sum_{\alpha_1\alpha_2} \sigma^{(2)}_{\ell \alpha_1\alpha_2} {\cal E}^{\rm in}_{1,\alpha_1} {\cal E}^{\rm in}_{2,\alpha_2}$. In the scalar potential gauge, the $\alpha$-th component of the electric field reads as following: ${\cal E}_{i,\alpha} = (q_{\alpha}/q) {\cal E}_i$, where ${\cal E}_i=\sqrt{ {\cal E}^2_{i,x}+{\cal E}^2_{i,y}}$ is the field amplitude. We conclude that the output electrical signal amplitude is proportional to the longitudinal component of the second-order conductivity, i.e.~${\cal E}^{\rm out} \propto \sigma^{(2)}_{\rm L} {\cal E}^{\rm in}_1 {\cal E}^{\rm in}_2$. Consequently, the conversion efficiency $\Upsilon_{\rm SF(DF)}$ for the sum- and difference-frequency (SF and DF, respectively) wave mixing processes can be estimated as following:
\begin{equation}\label{eq:efficiency-SF-DF}
\Upsilon_{\rm SF(DF)}= \frac{ I_{\rm SF(DF)} } {\sqrt{I_{1} I_{2}}} \approx
\frac {\sqrt{I_{1} I_{2}}}{8 n_{\omega_1} n_{\omega_2} n_{\omega_\Sigma} \epsilon^3_0 c^3} \big |\sigma^{(2),{\rm ee}}_{\rm L} \big |^2~,
\end{equation}
where $I_{i=1,2}$ indicates the intensities of the incoming laser beams and $I_{\rm SF(DF)}$ corresponds to the SF/DF-generated signal intensities.
Using Eqs. (\ref{eq:d2L1}) and  (\ref{eq:d2L2}), we find the following long-wavelength approximation for the longitudinal second-order conductivity:
\begin{widetext}
\begin{eqnarray}\label{eq:sigma2Leff}
&&\sigma^{(2),{\rm ee}}_{\rm L} (-{\bm q}_\Sigma, {\bm q}_1,{\bm q}_2,-\omega_\Sigma,\omega_1, \pm \omega_2)  =
 \frac{q_1}{{\cal R}_2} \Big \{  d^{(2)}_{xyyx}(-\omega_\Sigma;\omega_1, \pm \omega_2)
 \cos(\theta-\tilde\theta_1) \nonumber\\
 &+& \Big [d^{(2)}_{xyxy}(-\omega_\Sigma;\omega_1, \pm \omega_2) + 
d^{(2)}_{xyyx}(-\omega_\Sigma;\omega_1, \pm \omega_2)  \Big ]\cos(\theta+\tilde\theta_1) \Big \} \nonumber\\
&+& \frac{q_2}{{\cal R}_2} \Big \{  d^{(2)}_{xyyx}(-\omega_\Sigma; \pm \omega_2,\omega_1)\cos(\theta+\tilde\theta_2) 
+ \Big [d^{(2)}_{xyxy}(-\omega_\Sigma; \pm\omega_2,\omega_1) 
+d^{(2)}_{xyyx}(-\omega_\Sigma; \pm \omega_2,\omega_1)  \Big ]  \cos(\theta-\tilde\theta_2) \Big \}+\dots~.
\nonumber\\
\end{eqnarray}
\end{widetext}
Here, $\theta \equiv \theta_1-\theta_2$ is the relative angle, ${\tilde \theta}_i \equiv \theta_{\Sigma}-\theta_i$, and 
\begin{equation}\label{eq:tilde-theta}
\cos(\tilde\theta_i) = \frac{q^2_i \pm q_1q_2 \cos(\theta)}{q_i\sqrt{q^2_1+q^2_2 \pm 2q_1 q_2 \cos(\theta)}}~.
\end{equation}
In the SF and DF processes, the outgoing photon has frequency
\begin{equation}\label{eq:SF-DF-frequency}
\omega_\Sigma = \omega_1\pm \omega_2 \equiv \omega_{\rm SF(DF)}
\end{equation}
($\omega_{1}, \omega_{2} >0$) and wavevector
\begin{equation}\label{eq:SF-DF-wavevector}
|{\bm q}_{\Sigma}|=\sqrt{q^2_1+q^2_2 \pm 2 q_1 q_2 \cos(\theta)} \equiv |{\bm q}_{\rm SF(DF)}|~,
\end{equation}
where $\pm$ corresponds to the SF/DF case, respectively. By changing the value of the relative angle $\theta$, one can tune the strength of  $\sigma^{(2)}_{\rm L}$.

We start by analyzing the main features of the {\it bare} value of $\Upsilon_{\rm SF(DF)}$.
This is obtained by setting ${\cal R}_2 = 1$ in Eq.~(\ref{eq:sigma2Leff}) and inserting the result in Eq.~(\ref{eq:efficiency-SF-DF}). The corresponding numerical results are reported in Fig.~\ref{fig:YSFDFNof2}, where we plot $\Upsilon_{\rm SF(DF)}$ as a function of the relative frequency $\omega_{2} - \omega_{1}$, for a fixed value of $\omega_{1}$, and relative angle $\theta$. In the case of Fig.~\ref{fig:YSFDFNof2}, all the peaks that are seen originate from the intrinsic poles of the non-interacting second-order conductivity. We note that for $\omega_2=\omega_1$ there are sharp peaks in both $\Upsilon_{\rm SF}$ and $\Upsilon_{\rm DF}$. A large peak in $\Upsilon_{\rm SF}$ at $\omega_{1} = \omega_{2}$  (i.e.~$\omega_{\rm SF} = 2\omega$) indicates that giant SHG occurs in the quasi-homogeneous limit, which is remarkable since SHG is forbidden in graphene in the homogeneous limit due to its inversion symmetry\cite{Mikhailov_prb_2011}. On the other hand, a large peak in $\Upsilon_{\rm DF}$ at $\omega_{1} = \omega_{2}$ (i.e.~$\omega_{\rm DF} = 0$) indicates the finiteness of the ``photon drag'' (PD) response~\cite{Grinberg_prb_1988,Glazov_pr_2014}. Similarly to the case of SHG, optical rectification (which is, by definition, a phenomenon occurring in the $q=0$ limit) is forbidden in graphene because of its inversion symmetry. However, the finiteness of the second-order conductivity in the quasi-homogeneous limit enables the occurrence of the PD effect. Physically, PD means that we can induce a finite dc-current, in response to two laser beams, in a system with inversion symmetry thanks to {\it momentum transfer} from the photon to the electron subsystem.

We now proceed to analyze the role of electron-electron interactions. In Fig.~\ref{fig:YSFDFplasmon} we present numerical results, which, contrary to Fig.~\ref{fig:YSFDFNof2}, are now obtained by taking into account the factor $1/{\cal R}_2$ in Eq.~(\ref{eq:sigma2Leff}). In Fig.~\ref{fig:YSFDFplasmon}(a), we illustrate the functional dependence of $|{\cal R}_2|^{-1}$ on $\omega_{2} - \omega_{1}$ and $\theta$, for the case of the DF wave mixing process. We have discovered that ${\cal R}_2\sim1$ for the SF process, for the same parameters as in Fig.~\ref{fig:YSFDFplasmon}(a). Therefore, from now on, we concentrate only on the DF wave mixing process.

In agreement with Refs.~\onlinecite{Yao_prl_2014,Constant_np_2016}, we find that the DF wave mixing process is very effective to launch Dirac plasmons in graphene without the aid of a sharp AFM tip~\cite{fei_nature_2012,chen_nature_2012,woessner_naturemater_2015,alonsogonzalez_naturenano_2016}. This is because one can achieve frequency- and wavevector-matching between the outgoing photon generated in a DF wave mixing process and the Dirac plasmon. To this end, one needs to design the experiment in such a way to have sufficiently low-energy outgoing photons but with an in-plane wavevector which is much larger than that of incident photons. Dirac plasmon launching is therefore likely to occur in the DF wave mixing process~\cite{Yao_prl_2014,Constant_np_2016} because one can simply make $\omega_{\rm DF} = \omega_{\rm p}(|{\bm q}_{\rm DF}|)$ by changing the angle $\theta$. This  frequency-wavevector matching is more likely when $\theta$ approaches $\pi$. As we can see in Fig.~\ref{fig:YSFDFplasmon}, two plasmon resonances emerge for $\omega_{\rm DF}/\omega_1 <0.5$.

On the contrary, in the SF wave mixing process, the frequencies of the incoming lasers add up and result in an outgoing photon with higher frequency in comparison with that of the incident lasers. In the SF process one is not be able to fulfill the frequency-wavevector matching condition to launch low-energy Dirac plasmons. 

Electron-electron interactions alter also the PD effect in a significant manner.
The bare peak at $\omega_{2} = \omega_{1}$, which we highlighted while discussing Fig.~\ref{fig:YSFDFNof2}(b), disappears, while two plasmon resonances emerge. The former fact happens because the ${\cal R}^{-1}_{2}$ prefactor in Eq.~(\ref{eq:sigma2Leff}) is utterly small in the limit $\omega_{2} \to \omega_{1}$, for the case of the DF wave mixing process---see the dark region in Fig.~\ref{fig:YSFDFplasmon}(a). The reason for this behavior is easy to understand. In the DF wave mixing process, ${\cal R}_{2}$ is proportional to $1 + i q_{\rm DF} \sigma^{(1)}_{\rm L}(\omega_{\rm DF})/(2\epsilon_{0} \omega_{\rm DF})$. In the limit $\omega_{2} \to \omega_{1}$ ($\omega_{\rm DF} \to 0$), ${\rm Im}[\sigma^{(1)}_{\rm L}(\omega_{\rm DF})] =  4E_{\rm F} \sigma_{\rm uni}/(\pi \hbar \omega_{\rm DF})$ diverges like $1/\omega_{\rm DF}$ (Drude peak), leading to a divergence in ${\cal R}_{2}$. In other words, the bare PD efficiency is strongly diminished due to screening. In a real system with disorder, however, the $1/\omega_{\rm DF}$ divergence is regularized by a finite transport time (i.e.~the Drude peak is regularized into the low-frequency Drude tail). This implies that the PD efficiency in a disordered system is finite, contrary to the clean-limit case discussed above.

\begin{figure}[t]
\centering
\begin{overpic}[width=0.9\linewidth]{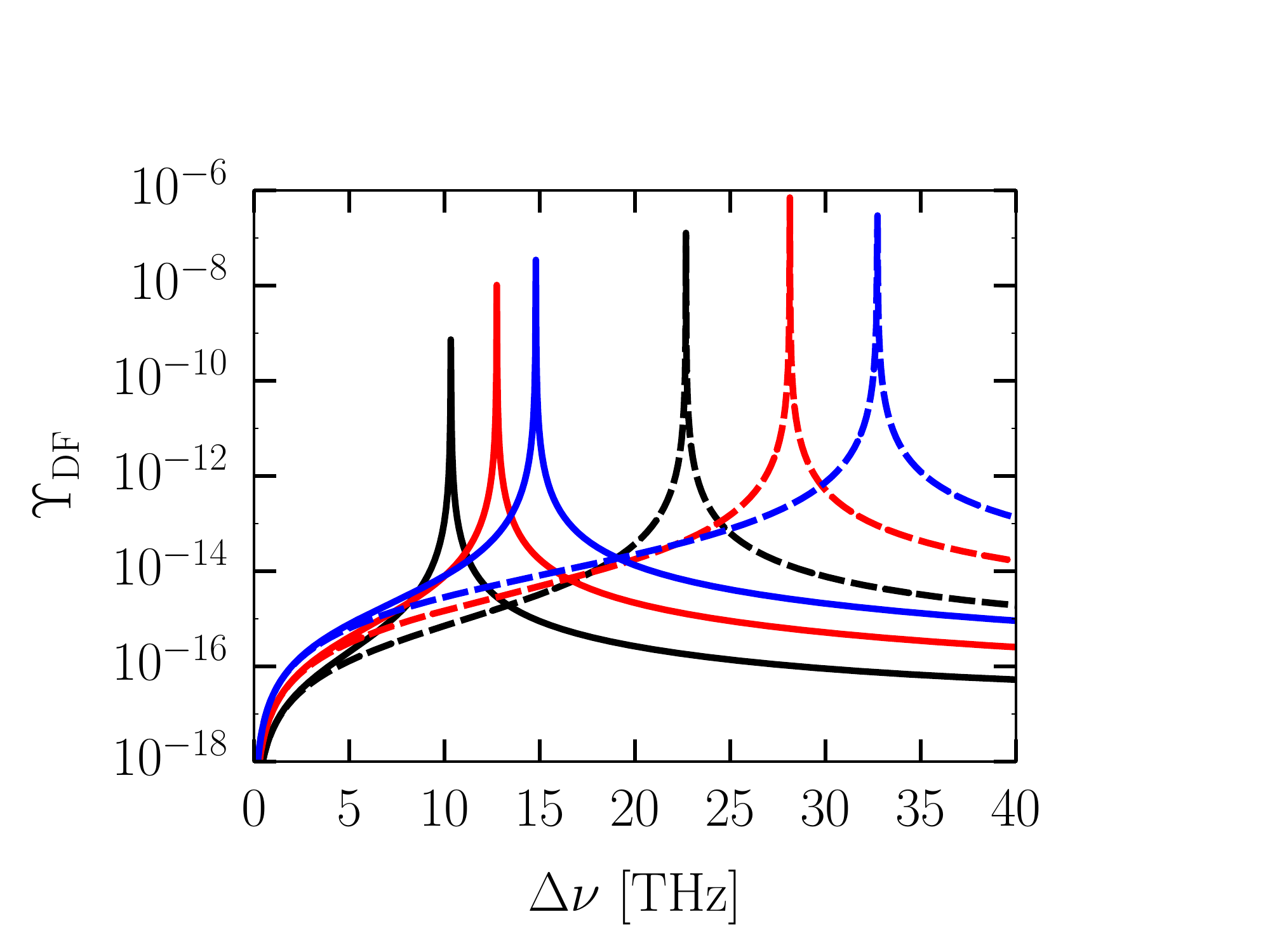}\put(-1,72){ (a)}\end{overpic}
\begin{overpic}[width=0.9\linewidth]{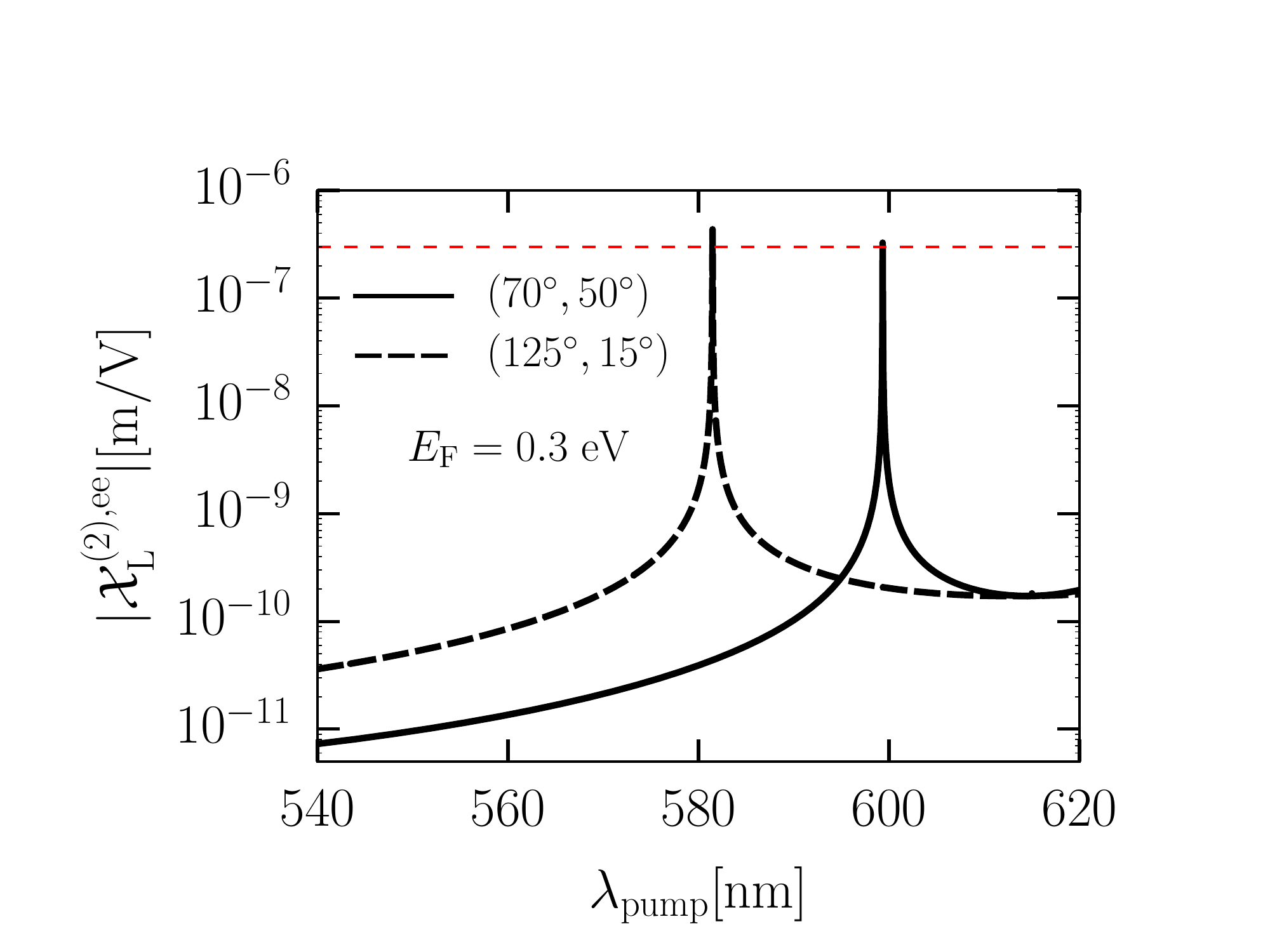}\put(-1,71){ (b)}\end{overpic}
\caption{(Color online) Panel (a) Theoretical predictions for the DF wave mixing conversion efficiency for two different laser configurations, as in Ref.~\onlinecite{Constant_np_2016}. Solid lines: $\vartheta_{\rm probe}=70^{\circ}$ and $\vartheta_{\rm pump}=50^{\circ}$. Dashed lines: $\vartheta_{\rm probe}=125^{\circ}$  and $\vartheta_{\rm pump}=15^{\circ}$. 
Different colors stand for different values of the Fermi energy $E_{\rm F}$: black, $E_{\rm F} = 200~{\rm meV}$, red: $300~{\rm meV}$, and blue: $400~{\rm meV}$. The pump beam wavelength is changed in the range $\lambda_{\rm pump}= 540$-$620~{\rm nm}$ in order to observe the plasmonic resonance for the different-frequency wave mixing signal at a frequency $\Delta\nu=c \big |\lambda^{-1}_{\rm pump}-\lambda^{-1}_{\rm probe} \big |$.  Panel (b) Our result for the dressed second-order susceptibility 
${\cal X}^{(2),{\rm ee}}_{\rm L}$ for the DF wave mixing process---Eq.~(\ref{eq:X2L})---is plotted as a function of the pump wavelength and for the same laser configurations as in panel (a).
The red dashed horizontal line indicates the value reported in Ref.~\onlinecite{Constant_np_2016}. Other parameters~\cite{Constant_np_2016}: 
$I_{\rm pump} = 100 I_{\rm probe} = 2~{\rm GW/ cm^2}$. The probe wavelength is fixed at $\lambda_{\rm probe}=615~{\rm nm}$.\label{fig:YDF}}
\end{figure}
\subsection{Comparison with available experimental results~\cite{Constant_np_2016}}
\label{subsection:comparison-with-experimental-results}

In Fig.~\ref{fig:YDF}(a) we present the DF wave mixing conversion efficiency for two different configurations of pump ($\omega_{\rm pump}$, $\vartheta_{\rm pump}$, and $q_{\rm pump}=\omega_{\rm pump}|\cos(\vartheta_{\rm pump})|/c$) and probe ($\omega_{\rm probe}$, $\vartheta_{\rm probe}$, and $q_{\rm probe}=\omega_{\rm probe}|\cos(\vartheta_{\rm probe})|/c$) laser beams, as in Ref.~\onlinecite{Constant_np_2016}. 

The angle between the in-plane wavevectors of the two beams is
\begin{equation}\label{eq:angle}
\theta=\frac{\pi}{2} [1-{\rm sign}(\vartheta_{\rm pump}-\pi/2) {\rm sign}(\vartheta_{\rm probe}-\pi/2)]~.
\end{equation}
In Ref.~\onlinecite{Constant_np_2016}, the authors explored two different angular configurations: $\vartheta_{\rm probe}=70^{\circ}$, $\vartheta_{\rm pump}=50^{\circ}$, and $\vartheta_{\rm probe}=125^{\circ}$, $\vartheta_{\rm pump}=15^{\circ}$. These two cases are illustrated in Fig.~\ref{fig:YDF}(a) by means of solid and dashed curves, respectively. Using Eq.~(\ref{eq:angle}), we therefore find $\theta=0$ and $\theta = \pi$ for the solid and dashed curves in Fig.~\ref{fig:YDF}(a), respectively. Sharp plasmon resonances are seen at the 
frequency difference between the two incident beams. As expected, the peaks display a blue-shift as Fermi energy increases. Moreover, we note that the curves corresponding to a relative angle $\theta=\pi$ (dashed curves) show stronger resonances at a higher frequency difference, in comparison with the solid ones. This is because this value of $\theta$ enables to launch Dirac plasmons with a larger wavevector (and therefore higher energy), a process that occurs with a large DF wave mixing conversion efficiency, as shown in Fig.~\ref{fig:YSFDFplasmon}(b). 

Dashed curves in Fig.~\ref{fig:YDF}(a) indicate plasmon launching in the frequency range $23$-$27~{\rm THz}$, in surprisingly good agreement with the experimental observation~\cite{Constant_np_2016} (i.e.~$\sim 23.8~{\rm THz}$ at $E_{\rm F}\sim 300~{\rm meV}$), considering that our theory does not deal with substrate effects and refers to a free-standing graphene sheet. Since our analysis is carried out in the case of a clean graphene sheet at zero temperature, the value of the efficiency $\Upsilon_{\rm DF}$ at the plasmon resonance is formally divergent and therefore impossible to be compared with the experimentally estimated dimensionless efficiency~\cite{Constant_np_2016} $\sim6\times 10^{-6}$. We, however, note that 
these formal mathematical divergencies are effectively regularized in the numerical calculations by i) the discrete nature of the wavelength mesh in the plots
and ii) the finite value of $\eta$, where $\eta$ has been defined in Sect.~\ref{sect:second-order-Dirac}, right after Eq.~(\ref{eq:form-factor}). Regarding point i), the numerical results in Fig.~\ref{fig:YDF} have been obtained by using a step $\delta \lambda_{\rm pump} =0.0123~{\rm nm}$ in the wavelength mesh. Regarding point ii), the results reported in Fig.~\ref{fig:YDF} (and all other figures above) have been obtained by setting $\eta/E_{\rm F}= 10^{-5}$. 

We now observe that the authors of Ref.~\onlinecite{Constant_np_2016} interpreted their experimental data by introducing a phenomenological model, where a second-order susceptibility, denoted by 
the symbol ${\cal X}^{(2)}$, was taken to be {\it frequency- and wavevector-independent}. {\it Fitting} their experimental data, the authors obtained~\cite{Constant_np_2016} 
$|{\cal X}^{(2)}| \approx 3 \times 10^{-7}~ {\rm m}/{\rm V}$.  Our theory gives a formal meaning to the quantity ${\cal X}^{(2)}$, identifying it with the plasmon-enhanced second-order susceptibility of graphene,
\begin{widetext}
\begin{equation}
{\cal X}^{(2)} \equiv {\cal X}^{(2),{\rm ee}}_{\rm L}(-{\bm q}_{\rm DF}, {\bm q}_{\rm probe},-{\bm q}_{\rm pump}, -\omega_{\rm DF}, \omega_{\rm probe},-\omega_{\rm pump})~.
\end{equation}
\end{widetext}
Here, the $n$-th order nonlinear susceptibility ${\cal X}^{(n),{\rm ee}}_{\ell \alpha_1\dots \alpha_n}$ is defined by the polarization response to the electric field, i.e.
\begin{equation}\label{eq:polarization}
P^{(n)}_{\ell} = \epsilon_0 \sum_{\{\alpha_i\}}{\cal X}^{(n),{\rm ee}}_{\ell \alpha_1\dots \alpha_n} \Pi^n_{i=1} {\cal E}_{\alpha_i}~.
\end{equation}
In a 2D system, the longitudinal component ${\cal X}^{(n),{\rm ee}}_{\rm L}$ of the nonlinear susceptibility is related to the $n$-th order longitudinal conductivity by
\begin{equation}\label{eq:X2L}
{\cal X}^{(n),{\rm ee}}_{\rm L} = i \frac{ \sigma^{(n),{\rm ee}}_{\rm L}}{ \epsilon_0  \omega_{\Sigma} d}~.
\end{equation}
In Eq.~(\ref{eq:X2L}), $d$ is an effective width in the direction perpendicular to the 2D electron system (i.e.~the effective system thickness). Numerical results for ${\cal X}^{(2),{\rm ee}}_{\rm L}$ for the case of graphene are presented in Fig.~\ref{fig:YDF}(b), where we have set $d=1~{\rm\AA}$. The horizontal dashed line in Fig.~\ref{fig:YDF}(b) denotes the experimental value $|{\cal X}^{(2)}| \approx 3 \times 10^{-7}~ {\rm m}/{\rm V}$. 
We therefore conclude that, for the aforementioned step in the wavelength mesh and value of $\eta$, our prediction for ${\cal X}^{(2),{\rm ee}}_{\rm L}$ on resonance matches the experimental value. We have checked that the peak value of ${\cal X}^{(2),{\rm ee}}_{\rm L}$ in Fig.~\ref{fig:YDF}(b) increases by a factor $\simeq 4$ when one reduces $\delta \lambda_{\rm pump}$ by a factor $4$. (We have also checked that a further reduction in the value of $\eta$ by a factor $10$ is irrelevant.)

\section{Summary and Conclusions}
\label{sect:summary}

We have presented a large-$N$ diagrammatic theory of plasmon-dressed second- and third-order nonlinear density response functions. Our work therefore represents a natural extension of the conventional linear-response Bohm-Pines random phase approximation~\cite{Pines_and_Nozieres,Giuliani_and_Vignale} to the realm of nonlinear response functions. Our most important formal results can be found in Eqs.~(\ref{eq:chi2_RPA}), (\ref{eq:chi3_RPAa}), and (\ref{eq:chi3_RPAb}).

While our theory is completely general, we have presented a wealth of numerical results for a specific two-dimensional material with inversion symmetry, i.e.~graphene. Electrons in this material are modelled in the usual fashion, by using the massless Dirac fermion model~\cite{Katsnelson,kotov_rmp_2012}. Our most important numerical results have been summarized in Figs.~\ref{fig:d2}, \ref{fig:plasmonic_shg_thg}, \ref{fig:YSFDFNof2}, \ref{fig:YSFDFplasmon}, and~\ref{fig:YDF}. More precisely, we have quantified second- and third-harmonic generation, sum- and difference-frequency wave mixing, and photon drag effects. All second-order effects (second-harmonic generation, sum- and difference-frequency wave mixing, and photon drag effects) in the quasi-homogeneous limit turn out to be finite and largely enhanced by plasmonic effects.

Detailed comparisons between our theory and available experimental results~\cite{Constant_np_2016} are reported in Section~\ref{subsection:comparison-with-experimental-results}.

This work can be generalized in the several directions. On the one hand, once can include electron-electron interaction effects beyond large-$N$ theory. This is notoriously difficult, although the most important corrections that are needed to deal with excitonic effects in semiconductors can be captured by ladder-type diagrams. More simply, one can take into account substrate and disorder effects, which may play an important role in establishing a truly microscopic description of nonlinear optics experiments like the one in Ref.~\onlinecite{Constant_np_2016}.

Last but not least, we note that our formal theory can be generalized to calculate 
nonlinear optical properties of two- and three-dimensional electron systems hosting topological plasmon modes~\cite{song_pnas_2016,kumar_prb_2016}, such as gapped graphene~\cite{gorbachev_science_2014,sui_naturephys_2015,shimazaki_naturephys_2015,beconcini_prb_2016}, transition-metal dichalcogenides~\cite{mak_science_2014,lee_naturenano_2016,saynatjoki_arxiv_2016}, and Weyl semimetals~\cite{huang_naturecommun_2015,huang_prx_2015,lv_prx_2015,lv_naturephys_2015,xu_science_2015,ghimire_jpcd_2015,pellegrino_prb_2015,Yang_np_2015,Shekhar_np_2015,xu_science_2015_2,xu_science_adv_2015,xu_prl_2016,inoue_science_2016,chang_science_adv_2016,wu_arxiv_2016}.

\acknowledgments
This work was supported by Fondazione Istituto Italiano di Tecnologia, the European Union's Horizon 2020 research and innovation programme under grant agreement No.~696656 ``GrapheneCore1'',  and the ERC Advanced Grant 338957 FEMTO/NANO (M.I.K.).

\begin{widetext}
\newpage 
\appendix
\section{Time-dependent Hartree theory of second- and third-order nonlinear response functions} 
\label{app:TDH}

In this Appendix we derive Eqs.~(\ref{eq:chi2_RPA}), (\ref{eq:chi3_RPAa}), and~(\ref{eq:chi3_RPAb}) by using the so-called time-dependent Hartree approximation (TDHA)~\cite{Giuliani_and_Vignale}.
 
We define the $n$-th order density response function of an interacting electron system as the $n$-th order functional derivative of the charge density with respect to the external scalar potential. For the first-, second-, and third-order density response functions we have
\begin{equation}\label{eq:chi1_def}
\chi^{(1)}({\bm r}_1,{\bm r}_2,t_1,t_2) \equiv \frac{\delta n({\bm r}_1,t_1)}{\delta V_{\rm ext}({\bm r}_2,t_2)}~,
\end{equation}
\begin{eqnarray}\label{eq:chi2_def}
\chi^{(2)}({\bm r}_1,{\bm r}_2,{\bm r}_3,t_1,t_2,t_3) &\equiv&\frac{\delta^2 n({\bm r}_1,t_1)}{\delta V_{\rm ext}({\bm r}_2,t_2)\delta V_{\rm ext}({\bm r}_3,t_3)}\nonumber\\ 
&=& \sum'_{\cal P}\frac{\delta \chi^{(1)}({\bm r}_1,{\bm r}_2,t_1,t_2)}{\delta V_{\rm ext}({\bm r}_3,t_3)}=
\frac{1}{2} \left [ \frac{\delta \chi^{(1)}({\bm r}_1,{\bm r}_2,t_1,t_2)}{\delta V_{\rm ext}({\bm r}_3,t_3)}+\frac{\delta \chi^{(1)}({\bm r}_1,{\bm r}_3,t_1,t_3)}{\delta V_{\rm ext}({\bm r}_2,t_2)}\right ]~,
\end{eqnarray}
and
\begin{eqnarray}\label{eq:chi3_def}
\chi^{(3)}({\bm r}_1,{\bm r}_2,{\bm r}_3,{\bm r}_4,t_1,t_2,t_3,t_4) &\equiv& \frac{\delta^3 n({\bm r}_1,t_1)}{\delta V_{\rm ext}({\bm r}_2,t_2)\delta V_{\rm ext}({\bm r}_3,t_3)\delta V_{\rm ext}({\bm r}_4,t_4)}\nonumber\\
&=& \sum'_{\cal P}\frac{\delta \chi^{(2)}({\bm r}_1,{\bm r}_2,{\bm r}_3,t_1,t_2,t_3)}{\delta V_{\rm ext}({\bm r}_4,t_4)}
= \frac{1}{3} \left[\frac{\delta \chi^{(2)}({\bm r}_1,{\bm r}_2,{\bm r}_3,t_1,t_2,t_3)}{\delta V_{\rm ext}({\bm r}_4,t_4)}\right.\nonumber\\
&+&\left.\frac{\delta \chi^{(2)}({\bm r}_1,{\bm r}_2,{\bm r}_4,t_1,t_2,t_4)}{\delta V_{\rm ext}({\bm r}_3,t_3)}
+\frac{\delta \chi^{(2)}({\bm r}_1,{\bm r}_3,{\bm r}_4,t_1,t_3,t_4)}{\delta V_{\rm ext}({\bm r}_2,t_2)}
\right ]~.
\nonumber\\
\end{eqnarray}
These definitions imply that the response functions are fully symmetric with respect to permutations of the variables ${\bm r}_i, t_i$ for $i\ge 2$. 

In the spirit of time-dependent density functional theory~\cite{Giuliani_and_Vignale}, we can build effectively non-interacting (i.e.~Kohn-Sham) response functions by differentiating the time-dependent density with respect to a suitable effective potential. In the spirit of the TDHA, the effective potential is solely given by the Hartree self-consistent potential: 
\begin{equation}
V_{\rm H}({\bm r,t}) =  \frac{1}{4\pi\epsilon_0} \int d{\bm r}_1 \frac{n({\bm r_1},t)}{|{\bm r}-{\bm r}_1|}
\end{equation}
where $\epsilon_0$ is the vacuum permittivity. The resulting TDHA response functions are given by:
\begin{eqnarray}\label{eq:chi123_0_def}
\chi^{(1)}_0({\bm r}_1,{\bm r}_2,t_1,t_2) &=&\frac{\delta n({\bm r}_1,t_1)}{\delta V_{\rm H}({\bm r}_2,t_2)}~,
\end{eqnarray}
\begin{eqnarray}
\chi^{(2)}_0({\bm r}_1,{\bm r}_2,{\bm r}_3,t_1,t_2,t_3) &=&\frac{\delta^2 n({\bm r}_1,t_1)}{\delta V_{\rm H}({\bm r}_2,t_2)\delta V_{\rm H}({\bm r}_3,t_3)}
=\sum'_{\cal P}\frac{\delta \chi^{(1)}_0({\bm r}_1,{\bm r}_2,t_1,t_2)}{\delta V_{\rm H}({\bm r}_3,t_3)}~,
\end{eqnarray}
and
\begin{eqnarray}
\chi^{(3)}_0({\bm r}_1,{\bm r}_2,{\bm r}_3,{\bm r}_4,t_1,t_2,t_3,t_4) &=&\frac{\delta^3 n({\bm r}_1,t_1)}{\delta V_{\rm H}({\bm r}_2,t_2)\delta V_{\rm H}({\bm r}_3,t_3)\delta V_{\rm H}({\bm r}_4,t_4)}
=\sum'_{\cal P}\frac{\delta \chi^{(2)}_0({\bm r}_1,{\bm r}_2,{\bm r}_3,t_1,t_2,t_3)}{\delta V_{\rm H}({\bm r}_4,t_4)}~.
\end{eqnarray}

Below, we will be using the following very useful functional derivative:
\begin{equation}\label{eq:useful_functional_derivative}
\frac{\delta V_{\rm H}({\bm r}_1,t_1)}{\delta V_{\rm ext}({\bm r}_2,t_2)} = \tilde \delta({\bm r}_1-{\bm r}_2)\tilde \delta(t_1-t_2) + \int d{\bm r}_3 dt_3 \frac{\delta V_{\rm H}({\bm r}_1,t_1) }{\delta n({\bm r}_3,t_3)} \frac{\delta n({\bm r}_3,t_3)}{\delta V_{\rm ext}({\bm r}_2,t_2)}~,
\end{equation}
where $\tilde \delta ({\bm r} - {\bm r}^\prime)$ indicates the Dirac delta function. After simple algebraic manipulations, Eq.~(\ref{eq:useful_functional_derivative}) becomes
\begin{equation}\label{eq:useful_functional_derivative_last}
\frac{\delta V_{\rm H}({\bm r}_1,t_1)}{\delta V_{\rm ext}({\bm r}_2,t_2)} = \tilde \delta({\bm r}_1-{\bm r}_2) \tilde\delta(t_1-t_2) + \int d{\bm r}_3 dt_3  
 v({\bm r_1},{\bm r}_3,t_1,t_3)  \chi^{(1)}({\bm r}_3,{\bm r_2},t_3,t_2)~.
\end{equation}
We now note that
\begin{equation}
v({\bm r}_1,{\bm r}_2,t_1,t_2) = \frac{\delta V_{\rm H}({\bm r}_1,t_1)}{\delta n({\bm r}_2,t_2)}= 
 \frac{1}{4\pi\epsilon_0}  \frac{\tilde \delta(t_1-t_2)}{|{\bm r}_1-{\bm r}_2|}~.
\end{equation}
In Fourier transform with respect to space,
\begin{equation}
v({\bm q}_1,{\bm q}_2) =\frac{1}{4\pi \epsilon_0} \int  \frac{d{\bm r}_1 d{\bm r}_2}{|{\bm r}_1 - {\bm r}_2|}  e^{i{\bm q}_1\cdot {\bm r}_1} e^{i{\bm q}_2\cdot {\bm r}_2} 
= v_{\bm q} \tilde\delta({\bm q}_1+{\bm q}_2)
\end{equation}
where $v_{\bm q} = 1/(\epsilon_0|{\bm q}|^2)$ [$v_{\bm q} = 1/(2\epsilon_0|{\bm q}|)$] for three-dimensional [two-dimensional] systems.

For the sake of notational simplicity, we will use from now on the following shorthand: $1\equiv ({\bm r}_1,t_1)$, $2\equiv ({\bm r}_2,t_2)$, etc. For example, Eq.~(\ref{eq:useful_functional_derivative_last}) reads as following:
\begin{equation}
\frac{\delta V_{\rm H}(1)}{\delta V_{\rm ext}(2)} = \tilde \delta(1,2) + \int d3 ~ v(1,3)  \chi^{(1)}(3,2)~.
\end{equation}
\subsection{First-order density response function in the TDHA}

Using the definitions given in the previous Section, we can write the following relation for the first-order density response function:
\begin{equation}
\chi^{(1)}(1,2)=\frac{\delta n(1)}{\delta V_{\rm ext}(2)}= \int d3 \frac{\delta n(1)}{\delta V_{\rm H}(3)}\frac{\delta V_{\rm H}(3)}{\delta V_{\rm ext}(2)}
= \int d3  \chi^{(1)}_0(1,3) \tilde \delta(3,2) + \int d3 \int d4  ~\chi^{(1)}_0(1,3)  v(3,4) \chi^{(1)}(4,2)~,
\end{equation}
or, more explicitly,
\begin{equation}\label{eq:chi1}
\chi^{(1)}(1,2) =  \chi^{(1)}_0(1,2) + \int d3 \int d4 ~\chi^{(1)}_0(1,3)   v(3,4)  \chi^{(1)}(4,2)~.
\end{equation}
In a compact matrix-form notation, Eq.~(\ref{eq:chi1}) reads as following:
\begin{equation}
\chi^{(1)}= \chi^{(1)}_0+ \chi^{(1)}_0  v  \chi^{(1)}~.
\end{equation}

In Fourier transform, the well-known RPA equation~\cite{Pines_and_Nozieres,Giuliani_and_Vignale} (\ref{eq:chi1}) becomes
\begin{equation}\label{eq:linear-response-RPA-general}
\chi^{(1)}({\bm q},{\bm q}',-\omega,\omega) = \chi^{(1)}_0({\bm q},{\bm q}',-\omega,\omega)+ 
\sum_{{\bm q}''} \chi^{(1)}_0({\bm q},{\bm q}'',-\omega,\omega) v_{{\bm q}''} \chi^{(1)}(-{\bm q}'',{\bm q}',-\omega,\omega)~.
\end{equation}
For translationally-invariant systems, Eq.~(\ref{eq:linear-response-RPA-general}) simplifies to the following well-known RPA result~\cite{Pines_and_Nozieres,Giuliani_and_Vignale}:
\begin{equation}\label{eq:linear-response-RPA-translationally-invariant}
\chi^{(1)}(-{\bm q},{\bm q},-\omega,\omega)
= \frac{\chi^{(1)}_{0}(-{\bm q},{\bm q},-\omega,\omega)}{1 - v_{\bm q} \chi^{(1)}_{0}(-{\bm q},{\bm q},-\omega,\omega)}~.
\end{equation}
We also introduce the RPA dynamical screening function~\cite{Pines_and_Nozieres,Giuliani_and_Vignale}:
\begin{equation}\label{eq:epsilon1}
\frac{1}{\epsilon({\bm q},\omega)} \equiv 1+v_{\bm q} \chi^{(1)}(-{\bm q},{\bm q},-\omega,\omega)= \frac{1}{1 - v_{\bm q}\chi^{(1)}_{0}(-{\bm q},{\bm q},-\omega,\omega)}~.
\end{equation}
Using $\epsilon_{\rm RPA}$, we can rewrite Eq.~(\ref{eq:linear-response-RPA-translationally-invariant}) as following:
\begin{equation}
\chi^{(1)}(-{\bm q},{\bm q},-\omega,\omega)= \frac{\chi^{(1)}_{0}(-{\bm q},{\bm q},-\omega,\omega)}{\epsilon({\bm q},\omega)}~.
\end{equation}

\subsection{Second-order density response function in the TDHA}

We now proceed to derive explicit expressions for the {\it second-order} density response. In this case, we need to perform the functional derivative of the first-order response $\chi^{(1)}$ with respect to the external field, as indicated in the second line of Eq.~(\ref{eq:chi2_def}). Using the representation of $\chi^{(1)}(1,2)$ given in Eq.~(\ref{eq:chi1}) inside Eq.~(\ref{eq:chi2_def}), we find:
\begin{eqnarray}\label{eq:chi2_eq1}
\chi^{(2)}(1,2,3) &=& \sum'_{\cal P}\frac{\delta \left \{ \chi^{(1)}_0(1,2) + \int d4 \int d5~ \chi^{(1)}_0(1,4)  v(4,5) \chi^{(1)}(5,2) \right \} }{\delta V_{\rm ext}(3)}  
 \nonumber \\
&=& \sum'_{\cal P} \Bigg [ \frac{\delta\chi^{(1)}_0(1,2)}{\delta V_{\rm ext}(3)}  + \int d4 \int d5 \Bigg \{ 
\frac{\delta\chi^{(1)}_0(1,4)}{\delta V_{\rm ext}(3)} v(4,5) \chi^{(1)}(5,2) 
+
\chi^{(1)}_0(1,4)   v(4,5) \frac{\delta\chi^{(1)}(5,2) }{\delta V_{\rm ext}(3)}
\Bigg\} \Bigg ]~.
\end{eqnarray}
We now need to calculate
\begin{eqnarray}\label{eq:d_chi1_0_dv}
\frac{\delta\chi^{(1)}_0(1,2)}{\delta V_{\rm ext}(3)} &=& \int d4~ \frac{\delta\chi^{(1)}_0(1,2)}{\delta V_{\rm H}(4)} \frac{\delta V_{\rm H}(4)}{\delta V_{\rm ext}(3)}
=\int d4 ~\chi^{(2)}_0(1,2,4) \left [ \tilde\delta(4,3) + \int d5~ v(4,5) \chi^{(1)}(5,3) \right ]
\nonumber\\  &=&
\chi^{(2)}_0(1,2,3)+ \int d4 \int d5 ~\chi^{(2)}_0(1,2,4) v(4,5)  \chi^{(1)}(5,3)~.
\end{eqnarray}
Using Eq.~(\ref{eq:d_chi1_0_dv}) in Eq.~(\ref{eq:chi2_eq1}), together with the definitions given in 
Eqs.~(\ref{eq:chi1_def}), (\ref{eq:chi2_def}), and~(\ref{eq:chi3_def}), we finally obtain:
\begin{eqnarray}\label{eq:second-order-intermediate}
\chi^{(2)}(1,2,3) &=& \chi^{(2)}_0(1,2,3)+  \int d4 \int d5 ~\chi^{(2)}_0(1,2,4)  v (4,5)  \chi^{(1)}(5,3)  
\nonumber\\ 
&+& \int d4 \int d5  ~\chi^{(2)}_0(1,4,3) v (4,5) \chi^{(1)}(5,2) 
\nonumber\\ 
&+&  \int d4 \int d5 \int d6 \int d7 ~\chi^{(2)}_0(1,4,5)  v (5,6) \chi^{(1)}(6,3)  v (4,7) \chi^{(1)}(7,2) 
\nonumber\\
&+& \int d4 \int d5 ~ \chi^{(1)}_0(1,4) v (4,5)  \chi^{(2)}(5,2,3)~.
\end{eqnarray}
Since the above result is symmetric with respect to the interchange of labels ``2'' and ``3'', 
we have dropped the permutation sign, i.e.~$\sum'_{\cal P}$.
We can rewrite Eq.~(\ref{eq:second-order-intermediate}) as
\begin{eqnarray}
&&\sum'_{\cal P}\int d5 \left[ \tilde\delta(1,5)  - \int d4 ~\chi^{(1)}_0(1,4) v (4,5) \right ] \chi^{(2)}(5,2,3) 
= \nonumber \\
&&\sum'_{\cal P} \int d4 \int d5 ~ \chi^{(2)}_0(1,4,5) 
\left [\tilde \delta(5,3)+ \int d6~  v (5,6) \chi^{(1)}(6,3)\right ] 
\left [ \tilde \delta(4,2)+ \int d7~ v (4,7)  \chi^{(1)}(7,2) \right ]~.
\end{eqnarray}
Using the following matrix-form notation  
\begin{eqnarray}
\tilde\delta(1,2)  - \int d3 ~\chi^{(1)}_0(1,3) v (3,2) & \equiv & 1-\chi^{(1)}_0 v
\nonumber\\ 
\tilde\delta(1,2)+ \int d3 ~ v (1,3) \chi^{(1)}(3,2) &\equiv& 1+ v~\chi^{(1)} 
\end{eqnarray}
and Eq.~(\ref{eq:epsilon1}), we can write the following compact expression for the second-order response function:
\begin{equation}
\chi^{(2)} =   \left [ 1+\chi^{(1)} v \right ]  \chi^{(2)}_0   \left [ 1+ v ~\chi^{(1)}  \right ] \left [ 1+v ~\chi^{(1)}  \right ]~.
\end{equation}
More explicitly,
\begin{eqnarray}\label{eq:chi2_eq2}
\chi^{(2)} (1,2,3)  &=&    \int d1'\int  d2' \int d3' \Bigg \{  
 \left [  \tilde\delta(1,1')+ \int d4 ~\chi^{(1)}(1,4) v(4,1') \right ]  \chi^{(2)}_0(1',2',3')  
\left [  \tilde\delta(3',3)+  \int d5 ~ v(3',5) \chi^{(1)}(5,3)  \right ]  \nonumber \\ &\times&
\left [ \tilde\delta(2',2)+ \int d6 ~ v(2',6) \chi^{(1)}(6,2)  \right ]  \Bigg \}~.
\end{eqnarray}
In Fourier transform and considering energy conservation, we find
\begin{eqnarray}
\chi^{(2)}({\bm q}_1,{\bm q}_2,{\bm q}_3,-\omega_\Sigma,\omega_2,\omega_3) &\equiv&  
\sum_{{\bm q}'_1,{\bm q}'_2,{\bm q}'_3} \left [ 1 +  \chi^{(1)}({\bm q}_1,{\bm q}'_1,-\omega_\Sigma,\omega_\Sigma) v_{{\bm q}'_1} \right ] 
 \chi^{(2)}_0(-{\bm q}'_1,-{\bm q}'_2,-{\bm q}'_3,-\omega_\Sigma,\omega_2,\omega_3) 
 \nonumber\\ &\times&  
\left [1+   v_{{\bm q}'_3} \chi^{(1)}({\bm q}'_3,{\bm q}_3,-\omega_3,\omega_3)  \right ]  
\left [1+   v_{{\bm q}'_2} \chi^{(1)}({\bm q}'_2,{\bm q}_2,-\omega_2,\omega_2)  \right ]~,
\end{eqnarray}
where $\omega_\Sigma=\omega_2+\omega_3$. 

In the case of translationally-invariant systems, the previous equation simplifies to
\begin{eqnarray}\label{eq:chi2-translationally-invariant}
\chi^{(2)}(-{\bm q}_\Sigma,{\bm q}_2,{\bm q}_3,-\omega_\Sigma,\omega_2,\omega_3) &=&  
\left [ 1 +  \chi^{(1)}(-{\bm q}_\Sigma,{\bm q}_\Sigma,-\omega_\Sigma,\omega_\Sigma) v_{{\bm q}_\Sigma} \right ] 
\chi^{(2)}_0(-{\bm q}_\Sigma,{\bm q}_2,{\bm q}_3,-\omega_\Sigma,\omega_2,\omega_3) 
\nonumber\\ &\times&  
\left [1+   v_{{\bm q}_3} \chi^{(1)}(-{\bm q}_3,{\bm q}_3,-\omega_3,\omega_3)\right ]  
\left [1+   v_{{\bm q}_2} \chi^{(1)}(-{\bm q}_2,{\bm q}_2,-\omega_2,\omega_2)\right ]~,
\end{eqnarray}
where ${\bm q}_\Sigma={\bm q}_2+{\bm q}_3$. Using Eq.~(\ref{eq:epsilon1}) inside Eq.~(\ref{eq:chi2-translationally-invariant}), we finally reach Eq.~(\ref{eq:chi2_RPA}) in the main text.

\subsection{Third-order density response function in the TDHA}

Finally, in this Section we derive an explicit expression for the third-order response in the TDHA. 
In this case, we need to take the functional derivative of the second-order response $\chi^{(2)}$ with respect to the external field, as in the second line of Eq.~(\ref{eq:chi3_def}). To this end, we first simplify the real-space representation of the second-order response:
\begin{eqnarray}\label{eq:chi2_eq3}
\chi^{(2)} (1,2,3)  
&=&  \chi^{(2)}_0(1,2,3)  
\nonumber\\
&+&  \int d1' \int d4'  ~ \chi^{(1)}(1,4') v(4',1')  \chi^{(2)}_0(1',2,3)  
\nonumber \\
&+&  \int d3' \int d5'  ~ \chi^{(2)}_0(1,2,3') v(3',5') \chi^{(1)}(5',3)
\nonumber \\
&+&  \int d2' \int d6' ~  \chi^{(2)}_0(1,2',3)  v(2',6') \chi^{(1)}(6',2)
\nonumber\\
&+& \int d1' \int d3' \int d4' \int d5' ~\chi^{(1)}(1,4') v(4',1')  \chi^{(2)}_0(1',2,3')  v(3',5') \chi^{(1)}(5',3)
\nonumber \\
&+& \int d1' \int d2' \int d4' \int d6'~ \chi^{(1)}(1,4') v(4',1')  \chi^{(2)}_0(1',2',3)  v(2',6') \chi^{(1)}(6',2)
\nonumber \\
&+& \int d2' \int d3' \int d5' \int d6' ~   \chi^{(2)}_0(1,2',3') v(3',5') \chi^{(1)}(5',3) v(2',6') \chi^{(1)}(6',2)
\nonumber \\
&+& \int d1' \int d2' \int d3' \int d4' \int d5' \int d6' ~  \chi^{(1)}(1,4') v(4',1') \chi^{(2)}_0(1',2',3') v(3',5') \chi^{(1)}(5',3) v(2',6') \chi^{(1)}(6',2)~.
\nonumber\\
\end{eqnarray}
Also, we need the following functional derivative:
\begin{eqnarray}\label{eq:dchi2_0_dv}
\frac{\delta \chi^{(2)}_0 (1,2,3) }{ \delta V_{\rm ext}(4)} = 
\int d5 \frac{\delta \chi^{(2)}_0(1,2,3) }{\delta V_{\rm H}(5)} \frac{\delta V_{\rm H}(5)}{\delta V_{\rm ext}(4)}
=\int d5 ~\chi^{(3)}_0(1,2,3,5)\left [ \tilde\delta(5,4) + \int d6 ~ v(5,6) \chi^{(1)}(6,4) \right ]~.
\end{eqnarray}
Using Eqs.~(\ref{eq:chi3_def}), (\ref{eq:chi2_eq3}), and~(\ref{eq:dchi2_0_dv}) and carrying out lengthy but straightforward algebra, we find:
\begin{eqnarray}\label{eq:chi3-full-TDHA}
\chi^{(3)}(1,2,3,4) &=& \sum'_{\cal P} \Bigg \{ 
 \int d5'~ \chi^{(3)}_0(1,2,3,5')\left [ \tilde\delta(5',4) + \int d6' ~ v(5',6') \chi^{(1)}(6',4) \right ]
\nonumber \\
&+&  \int d1' \int d4'  ~ \chi^{(2)}(1,4',4) v(4',1')  \chi^{(2)}_0(1',2,3)  
\nonumber\\
&+& \int d1' \int d4' \int d5'  ~ \chi^{(1)}(1,4') v(4',1')  \chi^{(3)}_0(1',2,3,5')  \left [ \tilde\delta(5',4) + \int d6' ~ v(5',6') \chi^{(1)}(6',4) \right ]
\nonumber \\
&+&  \int d3' \int d4'  ~ \chi^{(2)}_0(1,2,3') v(3',4') \chi^{(2)}(4',3,4)
\nonumber\\
&+& \int d3' \int d4' \int d5'  ~ \chi^{(2)}_0(1,2,3',5')  \left [ \tilde\delta(5',4) + \int d6' ~ v(5',6') \chi^{(1)}(6',4) \right ] v(3',4') \chi^{(1)}(4',3)
\nonumber \\
&+&   \int d2' \int d4' ~  \chi^{(2)}_0(1,2',3)  v(2',4') \chi^{(2)}(4',2,4)
\nonumber\\
&+& \int d3' \int d4' \int d5'  ~ \chi^{(3)}_0(1,2',3,5') \left [ \tilde\delta(5',4) + \int d6' ~ v(5',6') \chi^{(1)}(6',4) \right ]  v(2',4') \chi^{(2)}(4',2) 
\nonumber \\
&+& \int d1' \int d3' \int d4' \int d5' ~ \chi^{(2)}(1,4',4) v(4',1')  \chi^{(2)}_0(1',2,3')  v(3',5') \chi^{(1)}(5',3) 
\nonumber\\
&+&\int d1' \int d3' \int d4' \int d5' ~\chi^{(1)}(1,4') v(4',1')  \chi^{(2)}_0(1',2,3')  v(3',5') \chi^{(2)}(5',3,4)
\nonumber \\ 
&+& \int d1' \int d3' \int d4' \int d5' \int d6' ~\chi^{(1)}(1,4') v(4',1')  \chi^{(3)}_0(1',2,3',6') \left [ \tilde\delta(6',4) + \int d7' ~ v(6',7') \chi^{(1)}(7',4) \right ]  
\nonumber \\ &\times& v(3',5') \chi^{(1)}(5',3)
\nonumber\\
&+& \int d1' \int d2' \int d4' \int d6'~ \chi^{(2)}(1,4',4) v(4',1')  \chi^{(2)}_0(1',2',3)  v(2',6') \chi^{(1)}(6',2)
\nonumber\\
&+& \int d1' \int d2' \int d4' \int d6'~ \chi^{(1)}(1,4') v(4',1')  \chi^{(2)}_0(1',2',3)  v(2',6') \chi^{(2)}(6',2,4)
\nonumber \\
&+& \int d1' \int d2' \int d4' \int d5' \int d6'~ \chi^{(1)}(1,4') v(4',1')  \chi^{(3)}_0(1',2',3,5') \left [ \tilde\delta(5',4) + \int d7' ~ v(5',7') \chi^{(1)}(7',4) \right ] 
\nonumber \\ &\times& v(2',6') \chi^{(1)}(6',2)
\nonumber \\
&+& \int d2' \int d3' \int d4' \int d5' ~   \chi^{(2)}_0(1,2',3') v(3',4') \chi^{(2)}(4',3,4) v(2',5') \chi^{(1)}(5',2)
\nonumber\\
&+& \int d2' \int d3' \int d4' \int d5' ~   \chi^{(2)}_0(1,2',3') v(3',4') \chi^{(1)}(4',3) v(2',5') \chi^{(2)}(5',2,4)
\nonumber \\
&+& \int d2' \int d3' \int d4' \int d5' \int d6' ~   \chi^{(3)}_0(1,2',3',6') \left [ \tilde\delta(6',4) + \int d7' ~v(6',7') \chi^{(1)}(7',4) \right ] v(3',4') \chi^{(1)}(4',3) 
\nonumber \\ &\times& v(2',5') \chi^{(1)}(5',2)
\nonumber\\
&+& \int d1' \int d2' \int d3' \int d4' \int d5' \int d6' ~  \chi^{(2)}(1,4',4) v(4',1') \chi^{(2)}_0(1',2',3') v(3',5') \chi^{(1)}(5',3) v(2',6') \chi^{(1)}(6',2)
\nonumber\\
&+& \int d1' \int d2' \int d3' \int d4' \int d5' \int d6' ~  \chi^{(1)}(1,4') v(4',1') \chi^{(2)}_0(1',2',3') v(3',5') \chi^{(2)}(5',3,4) v(2',6') \chi^{(1)}(6',2)
\nonumber\\
&+& \int d1' \int d2' \int d3' \int d4' \int d5' \int d6' ~  \chi^{(1)}(1,4') v(4',1') \chi^{(2)}_0(1',2',3') v(3',5') \chi^{(1)}(5',3) v(2',6') \chi^{(2)}(6',2,4)
\nonumber\\
&+& \int d1' \int d2' \int d3' \int d4' \int d5' \int d6' \int d7' ~  \chi^{(1)}(1,4') v(4',1') \chi^{(3)}_0(1',2',3',7') 
\nonumber \\ &\times& \left [ \tilde\delta(7',4) + \int d8' ~v(7',8') \chi^{(1)}(8',4) \right ]  v(3',5') \chi^{(1)}(5',3) v(2',6') \chi^{(1)}(6',2)
\Bigg \}~.
\end{eqnarray}
The previous expression contains terms that vanish in the limit $\chi^{(3)}_{0}=0$: the sum of all these terms will be denoted by the symbol $\chi^{(3)}_{\rm a}$. However, we also notice the presence of terms that do {\it not} contain the non-interacting third-order response function $\chi^{(3)}_{0}$. Rather, they contain the non-interacting second-order response $\chi^{(2)}_{0}$. The sum of all these terms will be denoted by the symbol $\chi^{(3)}_{\rm b}$. In the diagrammatic language, the former family is depicted in Fig.~\ref{fig:chi3_RPA}, while the latter is depicted in Fig.~\ref{fig:chi3_bis_RPA}. 

Using Eq.~(\ref{eq:chi3-full-TDHA}) and Fourier transforming, we find that the quantity $\chi^{(3)}_{\rm a}$ obeys the following equation:
\begin{eqnarray}\label{eq:chi31}
&&\chi^{(3)}_{\rm a}({\bm q}_1,{\bm q}_2,{\bm q}_3,{\bm q}_4,-\omega_\Sigma,\omega_2,\omega_3,\omega_4) = 
\sum'_{\cal P} \Bigg \{
 \sum_{{\bm q}'_4}\chi^{(3)}_0({\bm q}_1,{\bm q}_2,{\bm q}_3,{\bm q}'_4,-\omega_\Sigma,\omega_2,\omega_3,\omega_4) 
\left [1 +  v_{{\bm q}'_4}  \chi^{(1)}(-{\bm q}'_4,{\bm q}_4,-\omega_4, \omega_4 ) \right ]
\nonumber\\
&&+  \sum_{{\bm q}'_2 {\bm q}'_4}\chi^{(3)}_0({\bm q}_1,{\bm q}'_2,{\bm q}_3,{\bm q}'_4,-\omega_\Sigma,\omega_2,\omega_3,\omega_4) 
 \left [1 +   v_{{\bm q}'_4} \chi^{(1)}(-{\bm q}'_4,{\bm q}_4,-\omega_4,\omega_4)  \right ]  v_{{\bm q}'_2}  \chi^{(1)} (-{\bm q}'_2,{\bm q}_2,-\omega_2,\omega_2)
\nonumber\\ 
&&+  \sum_{{\bm q}'_3 {\bm q}'_4}\chi^{(3)}_0({\bm q}_1,{\bm q}_2,{\bm q}'_3,{\bm q}'_4,-\omega_\Sigma,\omega_2,\omega_3,\omega_4) 
 \left [1 +   v_{{\bm q}'_4}  \chi^{(1)}(-{\bm q}'_4,{\bm q}_4,-\omega_4,\omega_4)  \right ]  v_{{\bm q}'_3} \chi^{(1)}(-{\bm q}'_3,{\bm q}_3,-\omega_3,\omega_3)
\nonumber\\ 
&&+  \sum_{{\bm q}'_1 {\bm q}'_4}\chi^{(1)}({\bm q}_1,-{\bm q}'_1,-\omega_\Sigma, \omega_\Sigma) v_{{\bm q}'_1}
\chi^{(3)}_0({\bm q}'_1,{\bm q}_2,{\bm q}_3,{\bm q}'_4,-\omega_\Sigma,\omega_2,\omega_3,\omega_4) \left [1 +   v_{{\bm q}'_4} \chi^{(1)}(-{\bm q}'_4,{\bm q}_4,-\omega_4,\omega_4)  \right ]   
\nonumber \\ 
&&+  \sum_{{\bm q}'_1 {\bm q}'_3 {\bm q}'_4}\chi^{(1)}({\bm q}_1,-{\bm q}'_1,-\omega_\Sigma,\omega_\Sigma) v_{{\bm q}'_1}
\chi^{(3)}_0({\bm q}'_1,{\bm q}_2,{\bm q}'_3,{\bm q}'_4,-\omega_\Sigma,\omega_2,\omega_3,\omega_4)  
\left [1+  v_{{\bm q}'_4} \chi^{(1)}(-{\bm q}'_4,{\bm q}_4,-\omega_4,\omega_4)   \right ]  
\nonumber \\ &&\times
v_{{\bm q}'_3} \chi^{(1)}(-{\bm q}'_3,{\bm q}_3,-\omega_3,\omega_3) 
\nonumber\\
&&+ \sum_{{\bm q}'_1 {\bm q}'_2 {\bm q}'_4}\chi^{(1)}({\bm q}_1,-{\bm q}'_1,-\omega_\Sigma,\omega_\Sigma) v_{{\bm q}'_1}
\chi^{(3)}_0({\bm q}'_1,{\bm q}'_2,{\bm q}_3,{\bm q}'_4,-\omega_\Sigma,\omega_2,\omega_3,\omega_4)  
\left [1+  v_{{\bm q}'_4} \chi^{(1)}(-{\bm q}'_4,{\bm q}_4,-\omega_4,\omega_4)   \right ]  
\nonumber \\ &&\times
v_{{\bm q}'_2} \chi^{(1)}(-{\bm q}'_2,{\bm q}_2,-\omega_2,\omega_2) 
\nonumber\\
&&+ \sum_{{\bm q}'_2 {\bm q}'_3 {\bm q}'_4} \chi^{(3)}_0({\bm q}_1,{\bm q}'_2,{\bm q}'_3,{\bm q}'_4,-\omega_\Sigma,\omega_2,\omega_3,\omega_4)   
\left [1+  v_{{\bm q}'_4} \chi^{(1)}(-{\bm q}'_4,{\bm q}_4,-\omega_4,\omega_4)   \right ]  
v_{{\bm q}'_3} \chi^{(1)}(-{\bm q}'_3,{\bm q}_3,-\omega_3,\omega_3)   
\nonumber \\ &&\times
v_{{\bm q}'_2} \chi^{(1)}(-{\bm q}'_2,{\bm q}_2,-\omega_2,\omega_2) 
\nonumber\\
&&+ \sum_{{\bm q}'_1 {\bm q}'_2 {\bm q}'_3 {\bm q}'_4}\chi^{(1)}({\bm q}_1,-{\bm q}'_1,-\omega_\Sigma,\omega_\Sigma) v_{{\bm q}'_1}
\chi^{(3)}_0({\bm q}'_1,{\bm q}'_2,{\bm q}'_3,{\bm q}'_4,-\omega_\Sigma,\omega_2,\omega_3,\omega_4)  
\left [1+  v_{{\bm q}'_4} \chi^{(1)}(-{\bm q}'_4,{\bm q}_4,-\omega_4,\omega_4)   \right ]  
\nonumber \\ &&\times
v_{{\bm q}'_3} \chi^{(1)}(-{\bm q}'_3,{\bm q}_3,-\omega_3,\omega_3) 
v_{{\bm q}'_2} \chi^{(1)}(-{\bm q}'_2,{\bm q}_2,-\omega_2,\omega_2) 
\Bigg \}~,
\end{eqnarray}
while the quantity  $\chi^{(3)}_{\rm b}$ is given by
\begin{eqnarray}\label{eq:chi32}
&&\chi^{(3)}_{\rm b}({\bm q}_1,{\bm q}_2,{\bm q}_3,{\bm q}_4,-\omega_\Sigma,\omega_2,\omega_3,\omega_4) =
\sum'_{\cal P} \Bigg \{
\sum_{{\bm q}'_1}  \chi^{(2)}({\bm q}_1,-{\bm q}'_1,{\bm q}_4,-\omega_\Sigma,\omega_\Sigma-\omega_4,\omega_4)  
v_{{\bm q}'_1} \chi^{(2)}_{0}({\bm q}'_1,{\bm q}_2,{\bm q}_3,-\omega_\Sigma+\omega_4,\omega_2,\omega_3) 
 \nonumber\\
&&+
\sum_{{\bm q}'_3}\chi^{(2)}_{0}({\bm q}_1,{\bm q}_2,{\bm q}'_3,-\omega_\Sigma,\omega_2,\omega_\Sigma-\omega_2)   v_{{\bm q}'_3} \chi^{(2)}(-{\bm q}'_3,{\bm q}_3,{\bm q}_4,-\omega_\Sigma+\omega_2,\omega_3,\omega_4)   
\nonumber\\
&&+
\sum_{{\bm q}'_2}\chi^{(2)}_{0}({\bm q}_1,{\bm q}'_2,{\bm q}_3,-\omega_\Sigma,\omega_\Sigma-\omega_3,\omega_3)   
v_{{\bm q}'_2}  \chi^{(2)}(-{\bm q}'_2,{\bm q}_2,{\bm q}_4,-\omega_\Sigma+\omega_3,\omega_2,\omega_4)   
\nonumber\\
&&+\sum_{{\bm q}'_1 {\bm q}'_3}\chi^{(2)}({\bm q}_1,-{\bm q}'_1,{\bm q}_4,-\omega_\Sigma,\omega_\Sigma-\omega_4,\omega_4)     
v_{{\bm q}'_1}  \chi^{(2)}_{0}({\bm q}'_1,{\bm q}_2,{\bm q}'_3,-\omega_\Sigma+\omega_4,\omega_2,\omega_3)       
v_{{\bm q}'_3}  \chi^{(1)}(-{\bm q}'_3,{\bm q}_3,-\omega_3,\omega_3)
\nonumber\\
&&+\sum_{{\bm q}'_1 {\bm q}'_2}\chi^{(2)}({\bm q}_1,-{\bm q}'_1,{\bm q}_4,-\omega_\Sigma,\omega_\Sigma-\omega_4,\omega_4)     
v_{{\bm q}'_1}  \chi^{(2)}_{0}({\bm q}'_1,{\bm q}'_2,{\bm q}_3,-\omega_\Sigma+\omega_4,\omega_2,\omega_3)       
v_{{\bm q}'_2}  \chi^{(1)}(-{\bm q}'_2,{\bm q}_2,-\omega_2,\omega_2)
\nonumber\\
&&+\sum_{{\bm q}'_1 {\bm q}'_3}  \chi^{(1)}({\bm q}_1,-{\bm q}'_1,-\omega_\Sigma,\omega_\Sigma) v({\bm q}'_1)  
\chi^{(2)}_{0}({\bm q}'_1,{\bm q}_2,{\bm q}'_3,-\omega_\Sigma,\omega_2,\omega_\Sigma-\omega_2)     
v_{{\bm q}'_3}  \chi^{(2)}(-{\bm q}'_3,{\bm q}_3,{\bm q}_4,-\omega_\Sigma+\omega_2,\omega_3,\omega_4)       
\nonumber\\
&&+\sum_{{\bm q}'_1 {\bm q}'_2}  \chi^{(1)}({\bm q}_1,-{\bm q}'_1,-\omega_\Sigma,\omega_\Sigma) v({\bm q}'_1)  
\chi^{(2)}_{0}({\bm q}'_1,{\bm q}'_2,{\bm q}_3,-\omega_\Sigma,\omega_\Sigma-\omega_3,\omega_3)     
v_{{\bm q}'_2}  \chi^{(2)}(-{\bm q}'_2,{\bm q}_2,{\bm q}_4,-\omega_\Sigma+\omega_3,\omega_2,\omega_4)     
\nonumber\\
&&+\sum_{{\bm q}'_2 {\bm q}'_3}  \chi^{(2)}_{0}({\bm q}_1,{\bm q}'_2,{\bm q}'_3,-\omega_\Sigma,\omega_2,\omega_\Sigma-\omega_2)  
v_{{\bm q}'_3}  \chi^{(2)}(-{\bm q}'_3,{\bm q}_3,{\bm q}_4,-\omega_\Sigma+\omega_2,\omega_3,\omega_4)   
v_{{\bm q}'_2}  \chi^{(1)}(-{\bm q}'_2,{\bm q}_2,-\omega_2,\omega_2)   
\nonumber\\
&&+\sum_{{\bm q}'_2 {\bm q}'_3}  \chi^{(2)}_{0}({\bm q}_1,{\bm q}'_2,{\bm q}'_3,-\omega_\Sigma,\omega_\Sigma-\omega_3,\omega_3) 
v_{{\bm q}'_3} \chi^{(1)}(-{\bm q}'_3,{\bm q}_3,-\omega_3,\omega_3) 
v_{{\bm q}'_2}  \chi^{(2)}(-{\bm q}'_2,{\bm q}_2,{\bm q}_4,-\omega_\Sigma+\omega_3,\omega_2,\omega_4)     
\nonumber\\
&&+  \sum_{{\bm q}'_1{\bm q}'_2{\bm q}'_3} \chi^{(2)}({\bm q}_1,-{\bm q}'_1,{\bm q}_4,-\omega_\Sigma,\omega_\Sigma-\omega_4,\omega_4) 
  v_{{\bm q}'_1} \chi^{(2)}_{0}({\bm q}'_1,{\bm q}'_2,{\bm q}'_3,-\omega_\Sigma+\omega_4,\omega_2,\omega_3)  
  v_{{\bm q}'_3} \chi^{(1)}(-{\bm q}'_3,{\bm q}_3,-\omega_3,\omega_3)  
  \nonumber\\ &&\times
  v_{{\bm q}'_2}\chi^{(1)}(-{\bm q}'_2,{\bm q}_2,-\omega_2,\omega_2)   
\nonumber\\
&&+ \sum_{{\bm q}'_1{\bm q}'_2{\bm q}'_3} 
\chi^{(1)}({\bm q}_1,-{\bm q}'_1,-\omega_\Sigma,\omega_\Sigma)   
v_{{\bm q}'_1}  \chi^{(2)}_{0}({\bm q}'_1,{\bm q}'_2,{\bm q}'_3,-\omega_\Sigma,\omega_2,\omega_\Sigma-\omega_2) 
v_{{\bm q}'_3} \chi^{(2)}(-{\bm q}'_3,{\bm q}_3,{\bm q}_4,-\omega_\Sigma+\omega_2,\omega_3,\omega_4) 
\nonumber\\ &&\times
v_{{\bm q}'_2}  \chi^{(1)}(-{\bm q}'_2,{\bm q}_2,-\omega_2,\omega_2) 
\nonumber\\
&&+ \sum_{{\bm q}'_1{\bm q}'_2{\bm q}'_3} 
\chi^{(1)}({\bm q}_1,-{\bm q}'_1,-\omega_\Sigma,\omega_\Sigma)   
v_{{\bm q}'_1}  \chi^{(2)}_{0}({\bm q}'_1,{\bm q}'_2,{\bm q}'_3,-\omega_\Sigma,\omega_\Sigma-\omega_3,\omega_3) 
v_{{\bm q}'_3} \chi^{(1)}(-{\bm q}'_3,{\bm q}_3,-\omega_3,\omega_3) 
\nonumber\\ &&\times
v_{{\bm q}'_2} \chi^{(2)}(-{\bm q}'_2,{\bm q}_2,{\bm q}_4,-\omega_\Sigma+\omega_3,\omega_2,\omega_4) 
\Bigg \}~.
\end{eqnarray}
As usual, for translationally-invariant systems, we can use momentum conservation (i.e.~${\bm q}_\Sigma={\bm q}_2+{\bm q}_3+{\bm q}_4$) for each response function. After lengthy but straightforward algebra, one reaches Eqs.~(\ref{eq:chi3_RPAa}) and~(\ref{eq:chi3_RPAb}) in the main text.

\section{Details on the derivation of Eqs.~(\ref{eq:d2L1}) and~(\ref{eq:d2L2})}
\label{app:d2L}

Since the second-order dipole ${\bm d}^{(2)}$ is a rank-$4$ tensor, it obeys the same symmetries of the third-order conductivity as in Eqs.~(\ref{eq:sym3_1}) and (\ref{eq:sym3_2}). The following equations follow from this observation:
\begin{eqnarray}
d^{(2)}_{\rm L,1}(\omega_1,\omega_2) &=&
 \frac{d^{(2)}_{xxyy,1}(-\omega_\Sigma;\omega_1,\omega_2) +d^{(2)}_{xyyx,1} (-\omega_\Sigma;\omega_1,\omega_2) }{2}  
 \cos(2\theta_1-\theta_2-\theta_\Sigma) 
 \nonumber \\ &+&
 \frac{d^{(2)}_{xxyy,1}(-\omega_\Sigma;\omega_1,\omega_2)  +2d^{(2)}_{xyxy,1}(-\omega_\Sigma;\omega_1,\omega_2) 
 +d^{(2)}_{xyyx,1}(-\omega_\Sigma;\omega_1,\omega_2)  }{2} 
 \cos(\theta_\Sigma-\theta_2)
 \end{eqnarray}
and
\begin{eqnarray}
d^{(2)}_{\rm L,2} (\omega_1,\omega_2) &=& 
 \frac{d^{(2)}_{xyxy,2}(-\omega_\Sigma;\omega_1,\omega_2)   + d^{(2)}_{xyyx,2} (-\omega_\Sigma;\omega_1,\omega_2) }{2}  \cos(2\theta_2-\theta_1-\theta_\Sigma) 
 \nonumber \\ &+& 
 \frac{d^{(2)}_{xyxy,2}(-\omega_\Sigma;\omega_1,\omega_2)  + 2 d^{(2)}_{xxyy,2} (-\omega_\Sigma;\omega_1,\omega_2) 
 + d^{(2)}_{xyyx,2} (-\omega_\Sigma;\omega_1,\omega_2)  }{2} \cos(\theta_\Sigma-\theta_1)~, 
\end{eqnarray}
where $\theta_\Sigma$ is given by
\begin{equation}
\cos(\theta_\Sigma-\theta_1) = \frac{q_1+q_2 \cos(\theta_2-\theta_1)}{\sqrt{q^2_1+q^2_2+2q_1 q_2 \cos(\theta_2-\theta_1)}}~.
\end{equation}
According to the intrinsic permutation symmetry of the second-order conductivity, we have
\begin{eqnarray}
&&\sigma^{(2)}_{\ell \alpha_1 \alpha_2} (-{\bm q}_{\Sigma},{\bm q}_1,{\bm q}_2,-\omega_\Sigma,\omega_1,\omega_2)
 -\sigma^{(2)}_{\ell \alpha_2 \alpha_1} (-{\bm q}_{\Sigma},{\bm q}_2,{\bm q}_1,-\omega_\Sigma,\omega_2,\omega_1) =
\sum_\beta \Bigg \{
 q_{1,\beta}  
 \Big [ d^{(2)}_{\ell \alpha_1\alpha_2\beta,1}(-\omega_\Sigma;\omega_1,\omega_2) 
 \nonumber \\ &&
-d^{(2)}_{\ell \alpha_2 \alpha_1\beta,2}(-\omega_\Sigma;\omega_2,\omega_1) \Big ]
+ q_{2,\beta} 
\Big [d^{(2)}_{\ell \alpha_1\alpha_2\beta,2}(-\omega_\Sigma;\omega_1,\omega_2)
-d^{(2)}_{\ell \alpha_2 \alpha_1\beta,1}(-\omega_\Sigma;\omega_2,\omega_1) \Big ] \Bigg \} + \dots
\equiv 0~.
\end{eqnarray}
We therefore can conclude that $d^{(2)}_{\ell \alpha_1\alpha_2\beta,2}(-\omega_\Sigma;\omega_1,\omega_2)
=d^{(2)}_{\ell \alpha_2 \alpha_1\beta,1}(-\omega_\Sigma;\omega_2,\omega_1)$.  This implies the following result 
\begin{eqnarray}
d^{(2)}_{\rm L,2}(\omega_1,\omega_2) &=& 
 \frac{d^{(2)}_{xxyy,1}(-\omega_\Sigma;\omega_2,\omega_1) +d^{(2)}_{xyyx,1} (-\omega_\Sigma;\omega_2,\omega_1) }{2}  
 \cos(2\theta_2-\theta_1-\theta_\Sigma) 
 \nonumber \\ &+&
\frac{d^{(2)}_{xxyy,1}(-\omega_\Sigma;\omega_2,\omega_1)  +2d^{(2)}_{xyxy,1}(-\omega_\Sigma;\omega_2,\omega_1) 
 +d^{(2)}_{xyyx,1}(-\omega_\Sigma;\omega_2,\omega_1)  }{2}  \cos(\theta_\Sigma-\theta_1)~.
\end{eqnarray}
Using Eq.~(\ref{eq:d2_derv}) we can conclude $d^{(2)}_{xyyx,1}=d^{(2)}_{xxyy,1}$. 
Using the above equations, we obtain Eqs.~(\ref{eq:d2L1}) and~(\ref{eq:d2L2}) in the main text. 

\section{Details on the derivation of Eq.~(\ref{eq:d2xxxx})}
\label{app:d2xxxx}

Taking the derivative in Eq.~(\ref{eq:d2_derv}), performing the integral over the polar angle of the vector ${\bm  k}$, and summing over the band indices $\{\lambda_1,\lambda_2,\lambda_3\}$ in Eq.~(\ref{eq:pi2}), we arrive at the following expression for $d^{(2)}_{xxxx}$:
\begin{eqnarray}\label{eq:to-be-integrated}
d^{(2)}_{xxxx}(-\omega_\Sigma;\omega_1,\omega_2) &=& 
\kappa\int^{\infty}_{0} dE \left [n_{\rm F}(E)F_0(E)  - n_{\rm F}(-E) F_0(-E) \right ] 
\nonumber \\&+&
\kappa\int^{\infty}_{0}  dE  \left [ n'_{\rm F}(E) F_1(E)  -  n'_{\rm F}(-E) F_1(-E) \right ]
\nonumber \\&+&
\kappa\int^{\infty}_{0}  dE  \left [n''_{\rm F}(E) F_2(E)  -   n''_{\rm F}(-E) F_2(-E) \right ]
\nonumber \\&+&
\kappa\int^{\infty}_{0}  dE  \left [n'''_{\rm F}(E) F_3(E)  -  n'''_{\rm F}(-E) F_3(-E) \right ]~.
\end{eqnarray}
Here, $\kappa= {e^3 \hbar v^2_{\rm F}}/{2\pi}$, $n_{\rm F}(E)$ is the usual the Fermi-Dirac distribution function at finite temperature $T$ and chemical potential $\mu$,
\begin{equation}
n_{\rm F}(E) = \left\{\exp\left(\frac{E-\mu}{k_{\rm B} T}\right)+1\right\}^{-1}~,
\end{equation}
\begin{eqnarray}
F_0(E) &\equiv& \frac{1}{4 \hbar\omega_1 \hbar \omega_2} 
\Bigg [ 
\frac{3}{2 E^2} 
+\frac{2\hbar\omega_2}{(2E + \hbar\omega_2)^3} - \frac{2\hbar\omega_2}{(2E-\hbar\omega_2)^3}
+\frac{2\hbar\omega_1}{(2E + \hbar\omega_1)^3} - \frac{2\hbar\omega_1}{(2E-\hbar\omega_1)^3}
\nonumber \\
&-&\frac{(\hbar\omega_1)^2+3\hbar\omega_1 \hbar\omega_2+3(\hbar\omega_2)^2}{(2E+\hbar\omega_1)^2\hbar\omega_2\hbar\omega_\Sigma}
-\frac{(\hbar\omega_1)^2+3\hbar\omega_1\hbar\omega_2+3(\hbar\omega_2)^2}{(2E-\hbar\omega_1)^2\hbar\omega_2\hbar\omega_\Sigma}
-\frac{3(\hbar\omega_1)^2+3\hbar\omega_1\hbar\omega_2+(\hbar\omega_2)^2}{(2E+\hbar\omega_1)\hbar\omega_1\hbar\omega_\Sigma}
\nonumber\\
&-&\frac{3(\hbar\omega_1)^2+3\hbar\omega_1\hbar\omega_2+(\hbar\omega_2)^2}{(2E-\hbar\omega_1)^2\hbar\omega_1\hbar\omega_\Sigma}
+\frac{(\hbar\omega_\Sigma)^2}{\hbar\omega_1\hbar\omega_2 (2E+\hbar\omega_\Sigma)^2}
+\frac{(\hbar\omega_\Sigma)^2}{\hbar\omega_1\hbar\omega_2 (2E-\hbar\omega_\Sigma)^2}
\Bigg ]~,
\end{eqnarray}
\begin{eqnarray}
 \nonumber \\
F_1(E) &\equiv& \frac{1}{4 \hbar\omega_1\hbar\omega_2 }
\Bigg [ 
\frac{\hbar\omega_1+2\hbar\omega_2}{\hbar\omega_1 \hbar\omega_\Sigma}
-\frac{3}{2 E} 
+\frac{\hbar\omega_2}{(2E-\hbar\omega_2)^2} -\frac{\hbar\omega_2}{(2E+\hbar\omega_2)^2}
+\frac{3\hbar\omega_1}{2(2E-\hbar\omega_1)^2} 
- \frac{\hbar\omega_1}{2(2E+\hbar\omega_1)^2} 
\nonumber \\
&+&\frac{3}{2(2E+\hbar\omega_1)}
-\frac{(\hbar\omega_\Sigma)^2}{\hbar\omega_1\hbar\omega_2(2E-\hbar\omega_\Sigma)}
+\frac{2(\hbar\omega_1)^2+3\hbar\omega_1\hbar\omega_2+3(\hbar\omega_2)^2}{2\hbar\omega_2(2E-\hbar\omega_1)\hbar\omega_\Sigma}
+\frac{3(\hbar\omega_1)^2+3\hbar\omega_1\hbar\omega_2+(\hbar\omega_2)^2}{2\hbar\omega_1(2E-\hbar\omega_2)\hbar\omega_\Sigma}
\nonumber\\
&+&\frac{3(\hbar\omega_1)^2+3\hbar\omega_1\hbar\omega_2+(\hbar\omega_2)^2}{2\hbar\omega_1(2E+\hbar\omega_2)\hbar\omega_\Sigma}
\Bigg ]~,
\end{eqnarray}
\begin{equation}
F_2(E) \equiv \frac{1}{2 \hbar\omega_1\hbar\omega_2} 
\Bigg [
1+ \frac{3 E}{2 \hbar\omega_1} -\frac{\hbar\omega_1}{4(2E-\hbar\omega_1)} +\frac{\hbar\omega_2}{8(2E+\hbar\omega_2)} -\frac{\hbar\omega_2}{8(2E-\hbar\omega_2)}\Bigg ]~,
\end{equation}
and
\begin{equation}
F_3(E) \equiv \frac{3 E}{8\hbar\omega_1\hbar\omega_2}~.
\end{equation}
Integrating by parts Eq.~(\ref{eq:to-be-integrated}), we reach
\begin{eqnarray}
d^{(2)}_{xxxx}(-\omega_\Sigma;\omega_1,\omega_2) &=& \kappa\Bigg\{ \int^\infty_0 n_{\rm F}(E) F(E) dE - \int^0_{-\infty} n_{\rm F}(E) F(E) dE\nonumber\\
&+&\lim_{E \to -\infty} G(E)-2 \lim_{E \to 0} G(E) +\lim_{E \to \infty} G(E)\Bigg \}~,
\end{eqnarray}
where
\begin{eqnarray}
F(E) &\equiv& F_0(E) - F'_1(E) +F''_2(E) -F'''_3(E) \nonumber \\
&=&\frac{1}{4 (\hbar\omega_2)^2} \left \{
\frac{\omega_1}{\omega_\Sigma }
\left [\frac{1}{(2E-\hbar\omega_1)^2}
-
\frac{1}{(2E+\hbar\omega_1)^2}
\right ]
+ \left (\frac{\omega_\Sigma}{\omega_1  }\right)^2
\left [
\frac{1}{(2E+\hbar\omega_\Sigma)^2}
-
\frac{1}{ (2E-\hbar\omega_\Sigma)^2} \right ]
\right \}
\end{eqnarray}
and
\begin{equation}
G(E) \equiv n_{\rm F}(E) \left [ F_1(E) - F'_2(E) +F''_3(E) \right ] +n'_{\rm F}(E) \left [F_2(E) -F'_3(E) \right ] +n''_{\rm F}(E) F_3(E)~.
\end{equation}
At zero temperature, 
\begin{equation}
\lim_{E \to -\infty} G(E) = - \frac{2\hbar\omega_1+\hbar\omega_2}{4(\hbar\omega_1)^2 \hbar\omega_2\hbar\omega_\Sigma}~,
\end{equation}
\begin{equation}
\lim_{E \to 0} G(E) = 0~,
\end{equation}
and
\begin{equation}
\lim_{E \to \infty} G(E) =0~.
\end{equation}
Since $F(E)$ is an odd function of $E$, we conclude that
\begin{equation}\label{eq:d2xxxx_int}
d^{(2)}_{xxxx}(-\omega_\Sigma;\omega_1,\omega_2) = \kappa \int^{\infty}_{0} dE ~ \left [ n_{\rm F}(E)+n_{\rm F}(-E)\right ] F(E) - \kappa \frac{2\hbar\omega_1+\hbar\omega_2}{4(\hbar\omega_1)^2 \hbar\omega_2\hbar\omega_\Sigma}~.
\end{equation}
Carrying out the integration in Eq.~(\ref{eq:d2xxxx_int}), we finally get Eq.~(\ref{eq:d2xxxx}) in the main text.
\end{widetext}
\end{document}